\documentclass[manuscript,screen,nonacm]{acmart}
\AtBeginDocument{
  }

\renewcommand\footnotetextcopyrightpermission[1]{}
\usepackage{xspace}
\usepackage{enumitem}
\usepackage{pifont}
\usepackage{multirow}
\usepackage{color,colortbl}
\usepackage{listings}
\usepackage{xcolor}
\usepackage{longtable}

\usepackage{soul}
\definecolor{VeryLightGreen}{rgb}{0.82, 1.0, 0.74}
\definecolor{lightgray}{gray}{0.85}
\sethlcolor{VeryLightGreen}

\newcommand\lt[1]{{\lstinline+#1+}}
\renewcommand\t[1]{{\lstinline+#1+}}
\definecolor{dkgreen}{rgb}{0,0.5,0}
\definecolor{dkred}{rgb}{0.5,0,0}
\definecolor{gray}{rgb}{0.5,0.5,0.5}
\newcommand{\toolname}{Kaizen\xspace}

\usepackage{tcolorbox}
\usepackage{placeins}
\setlist{topsep=2pt, itemsep=1pt, parsep=0pt}

\lstdefinelanguage{XML}{
	morecomment=[s]{<!--}{-->},
	morestring=[b]",
	moredelim=[s][\color{blue}]{<}{>},
	stringstyle=\color{red},
	commentstyle=\color{gray},
}

\lstdefinestyle{xmlStyle}{
	language=XML,
	basicstyle=\ttfamily\small,
	breaklines=true,
	frame=single
}

\usepackage{algorithm}
\usepackage{algpseudocode}
\usepackage{balance}

\definecolor{clrOffload}{HTML}{FDDCDC}
\definecolor{clrMemory}{HTML}{D4F1EF}
\definecolor{clrParallel}{HTML}{FEF9CC}
\definecolor{clrCode}{HTML}{E8E5FD}
\definecolor{clrImpl}{HTML}{FDDEED}
\definecolor{clrPerf}{HTML}{D5F8EE}
\definecolor{clrMisc}{HTML}{D9ECFD}
\definecolor{clrHeader}{HTML}{1E2535}
\definecolor{clrHeaderFg}{HTML}{FFFFFF}

\usepackage{booktabs, array, xcolor, colortbl, multirow}
\usepackage{tikz}

\newtcolorbox{findingbox}{
  colback=green!10!white,
  colframe=black!75!black,
  boxrule=1pt,
  leftrule=3mm,
  arc=1mm
}
\newcolumntype{L}[1]{>{\raggedright\arraybackslash}p{#1}}

\begin{document}

\title{\toolname: Metamorphic Fuzzing and Differential Testing for LLM-Translated HPC Applications}

\author{Oscar Ludwig}
\affiliation{
  \institution{Oregon State University}
  \city{Corvallis}
  \state{Oregon}
  \country{USA}
}
\email{ludwigo@oregonstate.edu}

\author{Ninad Anklesaria}
\affiliation{
  \institution{Oregon State University}
  \city{Corvallis}
  \state{Oregon}
  \country{USA}
}
\email{anklesan@oregonstate.edu}

\author{Zheming Jin}
 \authornote{Work was completed while affiliated with Oak Ridge National Laboratory}
\affiliation{
  \institution{Oak Ridge National Laboratory}
  \city{Oak Ridge}
  \country{USA}
}
\email{zheming.jin@gmail.com}

\author{Swaroop Pophale}
\affiliation{
  \institution{Oak Ridge National Laboratory}
  \city{Oak Ridge}
  \country{USA}
}
\email{pophaless@ornl.gov}

\author{Kausar Moshood}
\authornote{Authors contributed equally to this research.}
\affiliation{
  \institution{Oregon State University}
  \city{Corvallis}
  \state{Oregon}
  \country{USA}
}
\email{moshoodk@oregonstate.edu}

\author{Christian J. DeVore}
\authornotemark[2]
\affiliation{
  \institution{Oregon State University}
  \city{Corvallis}
  \state{Oregon}
  \country{USA}
}
\email{devorech@oregonstate.edu}

\author{Brandon Gill}
\authornotemark[2]
\affiliation{
  \institution{Oregon State University}
  \city{Corvallis}
  \state{Oregon}
  \country{USA}
}
\email{gillb3@oregonstate.edu}

\author{Cassius Villareal}
\authornotemark[2]
\affiliation{
  \institution{Oregon State University}
  \city{Corvallis}
  \state{Oregon}
  \country{USA}
}
\email{villarec@oregonstate.edu}

\author{Keita Teranishi}
\affiliation{
  \institution{Oak Ridge National Laboratory}
  \city{Oak Ridge}
  \country{USA}
}
\email{teranishik@ornl.gov}

\author{Manish Motwani}
\correspondingauthor
\affiliation{
  \institution{Oregon State University}
  \city{Corvallis}
  \state{Oregon}
  \country{USA}
}
\email{motwanim@oregonstate.edu}

\renewcommand{\shortauthors}{Ludwig et al.}

\begin{abstract}
Large language models (LLMs) are increasingly used to port scientific codes across heterogeneous high-performance computing (HPC) programming models, such as translating CUDA to OpenMP, OpenACC, Kokkos or SYCL. However, current evaluations use compilation success, token-level similarity, or developer-written tests from static benchmarks, which cannot reliably ensure behavioral correctness. We present \toolname, a metamorphic fuzzing and differential testing framework for evaluating the correctness of LLM-translated HPC code. \toolname uses metamorphic fuzzing via source-code mutation to generate semantically equivalent programs, grammar-based input fuzzing to explore behavioral diversity, and differential testing to expose semantic divergences between original and translated applications that compile and pass developer-written tests yet produce incorrect scientific results.

We evaluate \toolname on CUDA-to-OpenMP translation of 16 scientific applications from seven domains using three fine-tuned LLMs at kernel-level and full-program granularity. Our evaluation reveals that (1) compilation success is a poor proxy for correctness; (2) LLM-translated programs exhibit systematic compile-time error patterns, with nine categories for kernel-level translation and 27 for full-program translation; (3) semantic errors that survive compilation are often input-dependent and require differential testing to expose; and (4) full-program translation is substantially harder than kernel-level translation. These findings highlight the need for correctness-oriented evaluation of LLM-assisted HPC code translations.
\end{abstract}


\maketitle

\vspace{-1.5ex}
\section{Introduction}
\vspace{-0.8ex}
After the demise of Moore's law and Dennard scaling, modern high-performance computing (HPC) platforms have evolved primarily by increasing parallelism through multicore CPUs, then manycore architectures, and now accelerator (GPUs) based systems.
These architectural shifts have driven continual evolution of programming models and their specifications to sustain performance gains from massive parallelism and emerging hardware features. Realizing these gains when porting scientific applications to accelerator-based systems requires careful translation and adaptation of legacy code-bases that have evolved over decades.
Recent advances in LLM- and AI-assisted translation~(e.g., CodeRosetta~\cite{coderosetta}, Fortran2CPP~\cite{fortran2cpp}, LASSI~\cite{lassi}, and UniPar~\cite{unipar}) and performance-portable programming models and abstractions such as OpenMP~\cite{openmp}, SYCL~\cite{sycl}, OpenACC ~\cite{openacc}, and Kokkos~\cite{Kokkos} can reduce developer effort by providing functional, portable code that is not tightly coupled with the underlying hardware.
Our developed ChatPORT~\cite{ChatPORT_OMP} raised CUDA$\to$OpenMP kernel translation correctness rates by 43.2\% over baseline LLMs, with the best model reaching 79\%, and a SYCL extension~\cite{ChatPORT_SYCL} achieves up to 81.7\% correctness.
However, testing and verification remain significant bottlenecks as both, the capabilities of the LLMs and the specifications of these portable programming models evolve.

Unlike the sizable literature on performance portability and code modernization~\cite{10495889} illustrated by the U.S. DOE Exascale Computing Project’s successful adaptation of approximately 70~software products and 30~applications to accelerator platforms, there is very little prior work that focuses specifically on testing and verification of LLM-based scientific code translators.
In practice, evaluation remains largely manual and ad hoc: teams depend on informal comparisons, project-specific scripts, and sparse unit tests~\cite{Kanewala2014}.
Recurring gaps include the lack of systematic methods for establishing behavioral equivalence across programming models, inconsistent treatment of numerical stability under changing floating-point behavior, and weak or absent integration of correctness verification into routine workflows.
These gaps are exacerbated by LLM-translated codes, as programs using different programming models can look and behave in substantially different ways while achieving the same computational goals.
Consequently, even when automated translation accelerates code generation, developers must still expend substantial manual effort to re-establish confidence in correctness and scientific validity.

Current evaluation practices for LLM-based code translation rely on compilation success, token-level similarity, and developer-written tests from static benchmarks on which LLMs are likely to have been trained. None of these can reliably ensure that translated code preserves the original program's computational behavior.
A program can compile and execute successfully on standard test inputs while harboring semantic errors that manifest only under specific input conditions.
This reliance on shallow proxies creates false confidence in translation quality, particularly for HPC scientific applications where numerical correctness is paramount.

We present \toolname, a framework that combines metamorphic testing, grammar-based fuzzing, and differential testing to evaluate the effectiveness of LLM-based HPC code translators on arbitrarily generated HPC scientific programs by measuring the behavioral correctness of the LLM-translated programs.
\toolname uses semantics-preserving source code mutations to generate diverse program variants as translation inputs, applies runtime grammar-based input fuzzing to systematically explore the behavioral space, and employs differential testing to expose semantic divergences between original and translated programs that compile and pass fixed test inputs yet produce incorrect scientific results.
Although \toolname is designed to be generic across HPC programming models, in this work we prototype and evaluate it for CUDA-to-OpenMP translation across 16~scientific applications from 7~domains using three top-performing LLMs of ChatPORT~\cite{ChatPORT_OMP}, a suite of fine-tuned LLMs for HPC code porting, at both kernel-level (\emph{kernels} are computationally intensive, self-contained functions of an HPC application that execute on parallel hardware) and full-program translation granularity.

We investigate four research questions:

\textbf{\emph{RQ1:} To what extent do compilation success and developer-written tests from static benchmarks predict semantic correctness in LLM-translated HPC applications?}

\textbf{\emph{Answer:}} The predictive power of compilation success and developer-written tests is application-dependent: neither proxy can reliably ensure semantic correctness. Programs that compile and pass developer-written tests can produce wrong outputs under inputs that those tests never exercise, revealing a substantial gap between shallow evaluation proxies and true behavioral correctness.
For example, \emph{winograd} and \emph{lif} achieve compilation rates as high as 0.94 and 1.00 yet correctness of 0.00, while \emph{background-subtract} and \emph{cross} achieve up to 1.00~compilability and correctness, confirming that only differential testing with diverse inputs can reliably distinguish correct from incorrect translations.

\textbf{\emph{RQ2:} What compile-time errors do LLM-based HPC code translators introduce when translating CUDA-to-OpenMP kernels?}

\textbf{\emph{Answer:}} Kernel-level translations exhibit nine compile-time error categories. Several overlap with general LLM translation failures reported in prior work, while \emph{Incorrect Loop Construct}, \emph{Unsupported Construct Usage}, and \emph{Invalid Construct Usage} are specific to the semantic gap between CUDA's hierarchical execution model and OpenMP's directive-based programming model.

\textbf{\emph{RQ3:} What semantic errors survive compilation in LLM-translated HPC applications, and how effective is \toolname in exposing them?}

\textbf{\emph{Answer:}} Semantic errors in LLM-translated HPC applications are input-dependent and elude fixed developer-written tests. \toolname's metamorphic fuzzing and differential testing systematically exposes six categories of semantic errors: \emph{Intermediate Variable Elimination}, \emph{Execution Model Assumption Transfer}, \emph{Loop Bound Error}, \emph{Missing Statement Fault}, \emph{Shared Memory Scope Mistranslation}, and \emph{Multi-dimensional Index Flattening}. These errors produce incorrect scientific results only under specific input conditions, confirming that behavioral testing with diverse inputs is necessary to detect them.

\textbf{\emph{RQ4:} How does translation granularity (kernel vs full-program) affect LLM translation success and error profiles?}

\textbf{\emph{Answer:}} Full-program translation is substantially harder than kernel-level translation. While kernel-level translation achieves up to 72\% correctness, full-program translation fails to compile entirely for $ChatPORT_{\text{CL\_13B}}$, with errors spanning 27~categories across \emph{offloading}, \emph{memory management}, \emph{parallelism}, \emph{code validity}, and \emph{compatibility}.
Kernel-level fine-tuning degrades full-program compilation and correctness for two of the three ChatPORT variants, suggesting over-specialization toward kernel-only output.

This paper makes the following contributions:

\begin{itemize}
\item \toolname, a metamorphic fuzzing and differential testing framework for correctness evaluation of LLM-based HPC code translators.
\item A comprehensive empirical study of CUDA-to-OpenMP translation across 16~scientific applications from 7~domains using three top-performing fine-tuned LLMs at both kernel-level and full-program granularity.
\item A two-level taxonomy of compile-time errors in LLM-based CUDA-to-OpenMP translation, comprising 9~categories for kernel-level translation and 27~categories for full-program translation across six error groups.
\item Empirical evidence that neither compilation success nor developer-written tests from static benchmarks can reliably ensure semantic correctness, and that metamorphic testing with grammar-based fuzzing and differential testing is necessary to expose input-dependent semantic errors in LLM-translated HPC programs.
\item A replication package containing all code and data to replicate the results presented in this paper available at \url{https://github.com/ANSWER-OSU/Kaizen-Replication-Package}.
\end{itemize}

The remainder of this paper is organized as follows.
Section~\ref{sec:background} provides background on HPC code porting and LLM-based translation, and
Section~\ref{sec:motivation} describes the primary goals and challenges we address in this work and motivates the need for correctness-oriented evaluation of LLM-translators through a real-world example.
Section~\ref{sec:approach} describes the \toolname framework. Section~\ref{sec:evaluation} presents our experimental evaluation. Section~\ref{sec:relatedwork} places our work in the context of related work.
Section~\ref{sec:discussion} discusses our findings while
Section~\ref{sec:threats} addresses threats to their validity, and Section~\ref{sec:conclusion} summarizes our contributions.

\vspace{-1.5ex}
\section{Background}
\vspace{-0.8ex}
\label{sec:background}

This section describes the background on translating high-performance computing~(HPC) applications and automated testing techniques that \toolname builds upon.

\vspace{-1ex}
\subsection{HPC Code Porting and LLM-Based Translation}
\vspace{-0.5ex}
\label{sec:background-porting}

High-performance computing applications increasingly rely on GPU accelerators to sustain performance gains as CPU clock speeds plateau.
CUDA~\cite{cuda} remains the dominant programming model for NVIDIA GPUs, but its hardware-specific nature limits portability across diverse GPU architectures.
Performance-portable alternatives such as OpenMP target offloading~\cite{openmp}, SYCL~\cite{sycl}, OpenACC~\cite{openacc}, and Kokkos~\cite{Kokkos} allow developers to write code that runs across different hardware without modification.
Porting existing CUDA codebases to these models is a high-priority activity in the HPC community, particularly as organizations seek to run scientific applications on heterogeneous systems from multiple vendors.

Manual porting is labor-intensive and error-prone. A CUDA kernel (computationally intensive, self-contained function that executes on parallel hardware) must have its thread hierarchy, memory management, and synchronization primitives translated to semantically equivalent constructs in the target programming model.
For large scientific codebases, this can involve months of developer effort. LLM-based translation has emerged as a promising approach to accelerating this process.
Tools such as CodeRosetta~\cite{coderosetta}, Fortran2CPP~\cite{fortran2cpp}, LASSI~\cite{lassi}, and UniPar~\cite{unipar} use pre-trained or fine-tuned LLMs to automate translation across programming models.
ChatPORT~\cite{ChatPORT_OMP}, developed by the authors, is a suite of fine-tuned LLMs specialized for HPC code porting that raised CUDA-to-OpenMP correctness rates by 43.2\% over baseline LLMs, with the best model reaching 79\%~correctness. A SYCL extension~\cite{ChatPORT_SYCL} achieves up to 81.7\%~correctness on kernel-level translations.

Despite these advances, evaluation of LLM-based translation remains limited.
Most approaches rely on three metrics: compilation success, token-level similarity~\cite{codebleu}, and developer-written tests from static benchmarks on which LLMs are likely to have been trained.
Compilation success verifies syntactic validity but says nothing about whether the translated program computes the same results as the original.
Token-level similarity measures surface-level textual resemblance, which does not correlate reliably with semantic equivalence.
None of these can reliably ensure that a program does not harbor subtle semantic errors that manifest only under specific input conditions.

\vspace{-1ex}
\subsection{Testing LLM-Translated Code}
\vspace{-0.5ex}
\label{sec:background-testing}

Detecting semantic errors in LLM-translated code requires testing techniques that go beyond compilation checks. Three techniques are central to \toolname: metamorphic testing, fuzzing, and differential testing.

\textbf{Metamorphic testing}~\cite{segura2016} detects bugs by exploiting metamorphic relations: properties that specify how program outputs should change when inputs are transformed in a known way.
If a transformation preserves program semantics, the output behavior should be preserved before and after the transformation.
In \toolname, we apply semantics-preserving source code mutations to generate diverse program variants as translation inputs. If an LLM correctly translates a program, it should correctly translate \emph{all} semantically equivalent variants.
A translation that fails on a variant but succeeds on the original reveals a robustness gap in the LLM's translation capability and a translation correctness failure.

\textbf{Fuzzing}~\cite{Liang2018} is an automated testing technique that generates diverse inputs to exercise program behavior beyond what fixed test cases cover.
Grammar-based fuzzers generate inputs that conform to a specified syntax or structure, ensuring validity while maximizing diversity.
\toolname uses grammar-based fuzzing at two levels: source code fuzzing to generate diverse program variants as translation inputs, and runtime input fuzzing to explore the behavioral space of translated programs under diverse execution parameters.

\textbf{Differential testing}~\cite{Evans2007} runs two implementations of the same program on identical inputs and flags discrepancies between their outputs.
It is particularly effective for detecting semantic errors in translated code, where the original and translated versions should produce identical results.
\toolname applies differential testing by executing the original CUDA program and its LLM-translated OpenMP counterpart on the same fuzzed inputs, comparing outputs using configurable error norms to account for floating-point precision differences.

Together, these three techniques form a correctness-oriented evaluation methodology that goes beyond shallow proxies. Rather than asking whether a translation compiles successfully, passes developer-written tests, or resembles the ground-truth code at the token level, \toolname asks whether it behaves correctly across a diverse range of inputs and program variants.

\vspace{-1.5ex}
\section{Motivation}
\vspace{-0.8ex}
\label{sec:motivation}

This section motivates our work by describing the primary goals and challenges we address along with an illustrative real-world example showing the need for \toolname.

\vspace{-1ex}
\subsection{Goals and Challenges}
\vspace{-0.5ex}
\label{subsec:goals}

The primary goal of this work is to provide a systematic methodology to verify that LLM-translated HPC software preserves semantic correctness. Shallow evaluation proxies such as compilation success, token-level similarity, and developer-written tests from static benchmarks on which LLMs are likely to have been trained cannot reliably ensure that translated code exhibits the same computational behavior as the original implementation. This is particularly critical in HPC and scientific computing domains where numerical correctness and reproducibility are paramount. Evaluating semantic correctness of LLM-generated translations presents several unique challenges:

\noindent\textbf{Challenge 1: Data Leakage and Memorization.}
LLMs are trained on massive code corpora that may include benchmark suites and open-source HPC applications. When evaluated on these same benchmarks, high translation accuracy may reflect memorization rather than genuine translation capability. This data leakage makes it difficult to assess whether an LLM can generalize to unseen code patterns.

\noindent\textbf{Challenge 2: Limited Test Coverage.}
Conventional testing approaches rely on fixed, developer-written test suites that provide limited coverage of the input space, and these tests often come from the same static benchmarks on which LLMs are likely to have been trained. For complex HPC programs with multiple parameters, array dimensions, and computational paths, static test cases cannot adequately exercise all possible execution behaviors. Subtle semantic bugs may remain undetected under typical test inputs.

\noindent\textbf{Challenge 3: Hallucination and Subtle Semantic Errors.}
LLMs can generate syntactically valid code that compiles successfully but contains semantic errors invisible to syntax-based validation. These errors may include incorrect memory access patterns, wrong computational logic, improper synchronization primitives, or platform-specific behaviors that only manifest under specific conditions.

\noindent\textbf{Challenge 4: Cross-Platform Behavioral Differences.}
HPC programming models such as CUDA, OpenMP, and SYCL have different execution models, memory hierarchies, and synchronization semantics. A translation may produce correct results on one platform but exhibit undefined behavior or produce incorrect outputs on another due to subtle differences in parallel execution semantics.

To address these challenges, \toolname employs a two-level fuzzing strategy combined with metamorphic testing and differential testing to provide rigorous, behavior-focused evaluation of LLM-based HPC code translators.

\vspace{-1ex}
\subsection{A Motivating Example}
\vspace{-0.5ex}
\label{sec:motivation-example}

While LLMs have shown promise in translating HPC code between different programming models, these translations can introduce subtle correctness bugs that manifest only under specific input conditions. We demonstrate this challenge through a real-world example of the \emph{Leaky Integrate-and-Fire Neuron Model} from the HeCBench benchmark suite~\cite{Zheming2023}.

The Leaky Integrate-and-Fire (LIF) model is a fundamental computational neuroscience model that simulates the electrical behavior of biological neurons~\cite{gerstner2014}. The model is widely used in brain simulation, neuromorphic computing, and spiking neural networks. The LIF benchmark from HeCBench implements this model in CUDA, simulating the membrane voltage dynamics and spike generation of multiple neurons over discrete timesteps.

\textbf{Program Inputs and Outputs:}
The LIF program takes three command-line arguments that define the simulation configuration:
\begin{enumerate}
	\item \texttt{neurons\_per\_item}: Number of neurons in each computational item
	\item \texttt{num\_items}: Number of items to process
	\item \texttt{num\_steps}: Number of simulation timesteps
\end{enumerate}

The total number of neurons simulated is \texttt{neurons\_per\_item} $\times$ \texttt{num\_items}, and each neuron is simulated for \texttt{num\_steps} timesteps with a timestep size of $dt = 0.1$. The program outputs include the spike values for each neuron at the final timestep, along with the average kernel execution time. Correctness is verified by comparing the GPU implementation against a reference CPU implementation using a tolerance of $10^{-3}$.

\textbf{The Translation Bug:}
We used  $ChatPORT_{CL\_13B}$ LLM to translate \toolname-generated CUDA variant of the original implementation to OpenMP for CPU execution. The LLM successfully converted the CUDA kernel to OpenMP target directives and maintained the overall program structure. However, the translation introduced a critical semantic error: a single line of code was omitted from the neuron state update logic.

\begin{figure}[!t]
	\centering
    \includegraphics[scale=0.45]{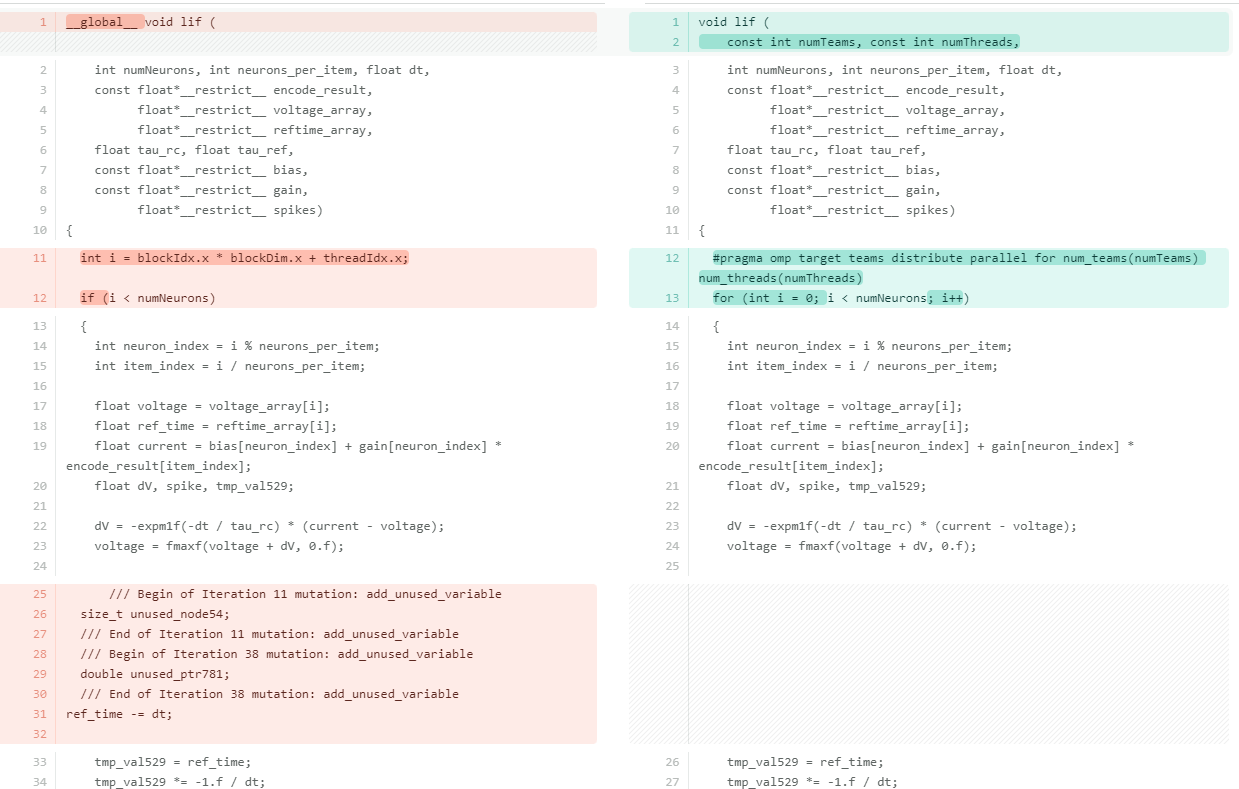}
	\caption{Partial diff of the CUDA-to-OpenMP translation of \toolname-modified variant of the Leaky Integrate-and-Fire (LIF) application from the HeCBench~\cite{Zheming2023} using  $ChatPORT_{CL\_13B}$. The CUDA kernel (left) contains \toolname-added unused variables that preserve program semantics. The translated OpenMP kernel (right) removed the unused variables along with the required refractory time decrement statement (line~31 on the left), breaking the neuron model's biological correctness.}
	\label{fig:lif_bug}
\end{figure}

Figure~\ref{fig:lif_bug} shows the diff between the \toolname-modified CUDA kernel (left) and the corresponding OpenMP-translated kernel (right).
As shown, \toolname randomly added two unused variables (lines~25--30) in the CUDA kernel before the original implementation's statement that decrements refractory time at each timestep (\texttt{ref\_time -= dt;}) (line~31). This statement implements the biological refractory period during which a neuron cannot fire again immediately after spiking.
While translating the kernel into OpenMP (right), the LLM intelligently removed the unused variable declaration statements and comments inserted by \toolname, but it also omitted the original code statement causing the refractory period mechanism to fail. Without proper refractory time updates, neurons can exhibit incorrect spiking behavior, producing scientifically invalid simulation results.

\textbf{Evading Developer-Provided Test Inputs:}
For each HeCBench application, we use developer-provided test inputs to start the fuzzing process.
Interestingly, for the LIF program, the buggy translation compiled and executed successfully on multiple developer-provided test inputs: \texttt{(100, 100, 10), (1000 100 10), (500 1000 50)}, each simulating distinct number of neurons for different timesteps. When we manually tested both the original CUDA and LLM-translated OpenMP versions with these test inputs, both programs produced identical outputs and passed verification.
A developer relying solely on these test inputs would incorrectly conclude that the translation is correct. This false negative occurs because the bug's manifestation is \textit{input-dependent}. The missing refractory time decrement only causes observable output differences when \emph{all} of the following three conditions are met.

\begin{enumerate}
	\item The random initialization produces specific neuron states for the given array configuration.
	\item Neurons enter refractory periods during the simulation.
	\item The simulation duration is sufficient for state divergence to accumulate.
\end{enumerate}

With the developer-provided test inputs, all these conditions are not met, allowing the buggy translation to produce outputs that coincidentally match the reference implementation.

\vspace{-1ex}
\subsection{\toolname's Systematic Bug Detection}
\vspace{-0.5ex}

\toolname employs grammar-based fuzzing and differential testing to systematically explore the input space beyond what developer-written tests cover.
\toolname generated diverse input configurations and successfully exposed the latent bug in the translated OpenMP version of the LIF program.
Table~\ref{tab:kaizen_results} shows two representative \toolname-generated inputs that revealed the incorrectness.

\begin{table}[!t]
	\centering
    \scriptsize
	\caption{\toolname-generated inputs expose the translation bug in LIF example, which are missed by the seed inputs.}
	\label{tab:kaizen_results}
	\begin{tabular}{llccc}
		\hline
		\textbf{Input type} & \textbf{Input value} & \textbf{CUDA} & \textbf{OpenMP} & \textbf{Outcome} \\
		\hline
		Seed & (100, 100, 10) & PASS & PASS & \textcolor{teal}{False negative} \\
		\hline
		Seed & (1000, 100, 10) & PASS & PASS & \textcolor{teal}{False negative} \\
		\hline
		Seed & (500, 1000, 50) & PASS & PASS & \textcolor{teal}{False negative} \\
		\hline
		Fuzzed & (10, 128, 1) & PASS & FAIL & \textcolor{red}{Bug detected} \\
		&  &  & @0: 0.000 vs 0.001 & \\
		\hline
		Fuzzed & (2048, 200, 200) & PASS & FAIL & \textcolor{red}{Bug detected} \\
		& & & @2192: 0.000 vs 0.001 & \\
		\hline
	\end{tabular}
\end{table}
\vspace{1ex}

\textbf{Fuzzed input 1: (10, 128, 1)} simulates 1,280 neurons for a single timestep. Despite the minimal configuration, the specific array size causes the random number generator to produce an initialization that exposes the bug immediately. The OpenMP version produces an incorrect spike value at neuron index 0.

\textbf{Fuzzed input 2: (2048, 200, 200)} simulates 409,600 neurons for 200 timesteps—a production-scale workload. The extended simulation duration amplifies the bug's effects as neurons spike and enter refractory periods throughout the execution. The mismatch occurs at neuron index 2192. Additionally, this input revealed a secondary issue: the OpenMP runtime emitted a warning about excessive thread requests, indicating the translation also failed to properly handle resource constraints.

These results demonstrate three critical insights that motivate \toolname's design.
\begin{enumerate}
	\item \textbf{Latent bugs evade manual testing}: The bug remains dormant under typical test inputs, creating false confidence in translation correctness.
	\item \textbf{Input-space exploration is essential}: Different configurations trigger different execution paths, exposing bugs that fixed inputs miss.
	\item \textbf{Semantic correctness requires differential testing}: Compilation success, developer-written tests, and runtime stability are insufficient; systematic output comparison across diverse inputs is necessary to ensure behavioral equivalence.
\end{enumerate}

\vspace{-1.5ex}
\section{The \toolname Approach}
\vspace{-0.8ex}
\label{sec:approach}

This section presents \toolname, a metamorphic fuzzing and differential testing framework for evaluating the behavioral correctness of LLM-translated HPC programs.
\toolname's primary goal is to provide a systematic methodology for evaluating semantic equivalence between source and translated code, going beyond compilation success, token-level similarity, and developer-written tests to verify that translations preserve computational behavior across diverse inputs and program variants.
We describe the overall architecture using CUDA$\rightarrow$OpenMP translation for illustration (Section~\ref{subsec:architecture}), followed by its two core components: source code metamorphic fuzzing (Section~\ref{subsec:source-fuzzing}) and runtime input fuzzing with differential testing (Section~\ref{subsec:runtime-fuzzing}).

\vspace{-1ex}
\subsection{\toolname Architecture}
\vspace{-0.5ex}
\label{subsec:architecture}

\begin{figure*}[!t]
	\centering
	\includegraphics[scale=0.5]{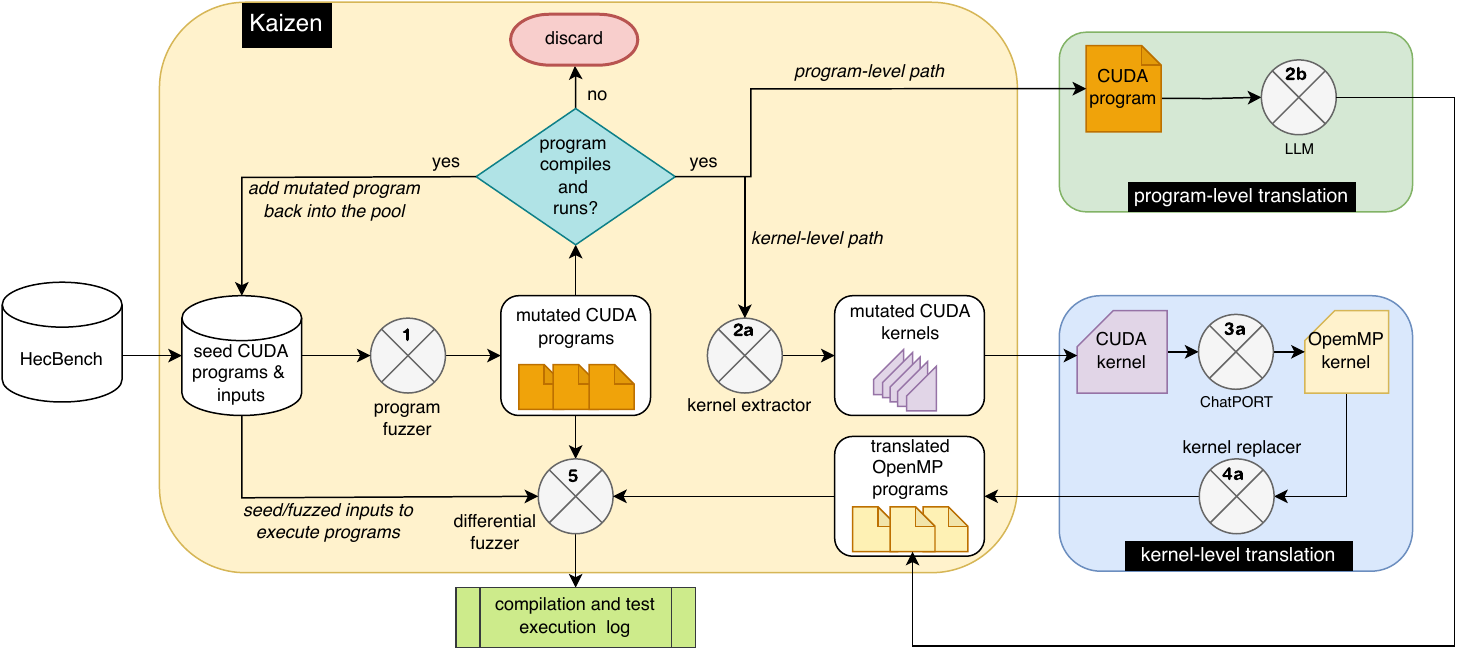}
	\caption{\toolname~framework architecture showing two parallel translation evaluation paths. The kernel-level path extracts and translates individual kernels (following ChatPORT's approach), while the program-level path translates complete CUDA applications. Both paths use source code fuzzing to generate diverse inputs and converge at differential testing for correctness validation.}
	\label{fig:architecture}
\end{figure*}

Figure~\ref{fig:architecture} illustrates the overall architecture of \toolname, which evaluates LLM-based code translation of HPC CUDA applications into OpenMP through two parallel pathways that both begin with source code fuzzing but differ in their translation scope.

\textbf{Stage 1: Source Code Metamorphic Fuzzing and Validation.}
Given original HPC CUDA applications, \toolname~applies a program-level fuzzer that uses the 15~grammar-based mutation operators (detailed in Table~\ref{tab:mutation-operators}) to generate diverse variants of CUDA programs. Each mutated program is validated by compiling and executing it with developer-provided inputs to ensure semantic preservation. Only variants that compile successfully and produce correct outputs proceed to the next stage, ensuring that downstream translation failures are attributable to the LLM rather than pre-existing bugs in the mutated source.
From this point, the framework branches into two parallel evaluation paths:

\vspace{-0.8ex}
\subsubsection{Kernel-Level Translation Path} This consists of Stages 2a--4a.

\textbf{Stage 2a: Kernel Extraction.}
For each validated CUDA program variant, \toolname~extracts individual computational kernels following ChatPORT's kernel-focused approach~\cite{ChatPORT_OMP, ChatPORT_SYCL}, isolating device-side computation from host-side orchestration code.
This extraction is necessary because ChatPORT accepts a single kernel file as input.

\textbf{Stage 3a: Kernel Translation.}
Each extracted CUDA kernel is translated using the three top-performing ChatPORT LLM variants: $ChatPORT_{\text{CL\_13B}}$, $ChatPORT_{\text{SCB\_7B}}$, and $ChatPORT_{\text{HPC\_C\_6.7B}}$~\cite{ChatPORT_OMP}.
These models were fine-tuned on CUDA-to-OpenMP kernel pairs from HeCBench and the OpenMP Validation and Verification suite~\cite{Diaz2018}. The framework collects all generated OpenMP kernel translations along with any compilation warnings or errors.

\textbf{Stage 4a: Kernel Replacement.}
Translated OpenMP kernels are inserted back into the original developer-written OpenMP programs provided in HeCBench, replacing existing human-written OpenMP kernels while keeping host-side code (memory management, data transfers, kernel launch configurations) unchanged. This design ensures that any correctness failures observed during differential testing can be attributed to kernel translation rather than host-side code issues. This step produces complete OpenMP programs ready for compilation and execution.

\begin{figure}[!t]
\raggedright
\begin{tcolorbox}
\small \textbf{LLM Prompt template}: You are an HPC developer with knowledge of CUDA and OpenMP programming models. Translate the following CUDA program into OpenMP while preserving the CUDA runtime semantics and replacing CUDA APIs with OpenMP counterparts. Only generate the program and do not provide any explanations or comments.
\\
\\
$<$CUDA source code$>$
\end{tcolorbox}
\vspace{-1ex}
\caption{Prompt template engineered after manually translating a sample of 5 HeCBench programs from CUDA to OpenMP using the top-3 performing LLMs (HPC\_Coder, StarCoder, and CodeLlama) for kernel-level translations in ChatPORT~\cite{ChatPORT_OMP}.}
\label{fig:full_translation_prompt}
\vspace{-2ex}
\end{figure}

\vspace{-0.8ex}
\subsubsection{Full-Program Translation Path.} This consists of Stage 2b.

\textbf{Stage 2b: Full Program Translation.}
Each validated CUDA program variant is fed in its entirety to LLM to assess its capabilities in translating complete programs.
The LLMs must translate not only the computational kernels but also all host-side code including memory allocation (\texttt{cudaMalloc}~$\to$~OpenMP target data mapping), data transfers (\texttt{cudaMemcpy}~$\to$~OpenMP target update), kernel launches (CUDA syntax~$\to$~OpenMP target teams distribute), and error handling. This represents a significantly more complex translation task than isolated kernel translation.

We use the same three ChatPORT variants used for kernel-level translation, since no models fine-tuned specifically for full-program translation are currently available.
To engineer an effective prompt for full-program translation, multiple authors independently and iteratively developed prompts on a held-out set of 5~HeCBench applications (\emph{Adam}, \emph{Concat}, \emph{Goulash}, \emph{Mrc}, and \emph{Overlay}), comparing LLM-translated versions against developer-written implementations.
Authors reconciled their prompts after each iteration, arriving at the prompt template shown in Figure~\ref{fig:full_translation_prompt}.

\vspace{-0.8ex}
\subsubsection{Convergence: Differential Testing} This consists of Stages 5a--5c:

\textbf{Stage 5a: Compilation Validation.}
Each translated complete program is compiled using the OpenMP-enabled compiler (e.g., AMD Clang/LLVM). Programs that fail compilation are recorded with their error messages for analysis under RQ2. Successfully compiled programs proceed to next stage.

\textbf{Stage 5b: Runtime Input Fuzzing.}
For each successfully compiled translation, \toolname~generates a comprehensive set of runtime test inputs by fuzzing kernel parameters, array dimensions, and input data values. This runtime fuzzing uses grammar-based input generation to systematically explore the behavioral space to maximize behavioral coverage.

\textbf{Stage 5c: Differential Execution and Comparison.}
The original CUDA program and the translated OpenMP program execute on the same fuzzed inputs. The original runs on NVIDIA GPUs while the translated version runs on CPUs with OpenMP offloading support.
\toolname~compares program outputs using configurable error norms and flags discrepancies where outputs exceed the error norms or result in runtime crashes or timeouts as correctness failures.

\vspace{-1ex}
\subsection{Source Code Metamorphic Fuzzing}
\vspace{-0.5ex}
\label{subsec:source-fuzzing}

The first component of \toolname~systematically generates diverse variants of input source code to evaluate LLM translation robustness.
Unlike random mutation approaches, \toolname employs grammar-aware transformations that preserve program semantics while maximizing syntactic diversity.
This approach applies to both kernel-level and full program translation scenarios.

\vspace{-0.8ex}
\subsubsection{Mutation Operators}
\vspace{-0.4ex}

Table~\ref{tab:mutation-operators} shows 15 carefully designed mutation operators that \toolname~employs to generate semantically equivalent code variants while introducing syntactic and structural diversity. These operators are organized into the following five categories based on their transformation strategy:

\begin{table}[!t]
	\centering
    \scriptsize
	\caption{Source code mutation operators in \toolname~for generating diverse translation inputs. All operators preserve semantic equivalence while introducing syntactic variations.}
	\label{tab:mutation-operators}
	\begin{tabular}{llp{8cm}}
		\toprule
		\textbf{ID} & \textbf{Mutation Operator} & \textbf{Description} \\
        \toprule
		M1 & Add dead if-statement block & Insert if-statement block with false conditional (e.g., \texttt{if (false)})  \\
        \hline
		M2 & Add dead switch-statement block & Insert switch-statement block with false or unreachable case labels \\
        \hline
		M3 & Add dead loop & Insert for/while loop with false conditional (e.g., \texttt{while (false)}) \\
        \hline
		M4 & Add unused variable & Declare and optionally initialize variables that are never used in the program \\
        \hline
	    M5 & Add comments & Insert natural language comments that may contradict actual code behavior \\
        \hline
	    M6 & Edit comments & Modify existing comments with alternatives \\
        \hline
	    M7 & Delete comments & Remove previously inserted comments \\
		\hline
	    M8 & Swap variables & Reorder consecutive variable declarations that have no data dependencies \\
        \hline
	    M9 & Reorder predicates & Change the order of predicates within if-statements or loop conditionals \\
        \hline
	    M10 & Add predicates & Insert conditional predicates that preserve behavior (e.g., add \texttt{|| false} or \texttt{\&\& true}) \\
		\hline
        M11 & Rename variable & Change variable name and update all subsequent uses throughout the scope \\
        \hline
	    M12 & Delete added if-statement & Remove previously inserted dead if-statement block \\
        \hline
	    M13 & Delete added switch block & Remove previously inserted dead switch-statement block \\
        \hline
	    M14 & Delete added loop & Remove previously inserted dead for/while loop and all code inside \\
        \hline
	    M15 & Delete added unused variables & Remove previously inserted unused variable declarations \\
	    \bottomrule
	\end{tabular}
\end{table}
\vspace{1ex}

\begin{enumerate}
    \item \textbf{Dead Code Injection Operators (M1--M4)} insert unreachable code that does not affect program execution, testing whether the LLM can distinguish executable logic from unreachable code and whether spurious code confuses the translation process.

    \item \textbf{Comment Manipulation Operators (M5--M7)} modify documentation to assess whether the LLM relies on comments versus actual code logic for translation decisions. By inserting misleading comments, editing existing ones, or removing them entirely, \toolname tests whether translation quality depends on documentation cues.

    \item \textbf{Structural Reordering Operators (M8--M10)} permute semantically independent code elements, testing whether translation depends on specific code orderings or whether the LLM correctly understands data flow independence.

    \item \textbf{Renaming Operator (M11)} systematically change variable names and their references within scope, testing whether translation quality depends on specific naming conventions or whether the LLM correctly analyzes program structure independent of identifier choices.

    \item \textbf{Deletion Operators (M12--M15)} remove previously inserted constructs, enabling the framework to explore both addition and removal of code constructs as reverse mutations.
\end{enumerate}

Each mutation is validated through compilation and testing using developer-provided tests to ensure semantic preservation before being used as a translation input.

\vspace{-0.8ex}
\subsubsection{Grammar-Based Source Code Mutation Strategy}
\vspace{-0.4ex}

\toolname~uses a grammar-based approach that respects the syntactic structure and semantic constraints of the source programming language.
For example, to evaluate C++-based implementations, we use C++ parsers and abstract syntax tree (AST) representations to ensure all generated variants remain well-formed and semantically equivalent.

The fuzzing strategy proceeds iteratively by starting from a seed program (e.g., original HeCBench CUDA application), \toolname~applies mutation operators at randomly selected AST nodes, validates semantic preservation through compilation and baseline testing against developer-provided test inputs and reference outputs, and adds valid variants to the corpus.
This process continues until reaching user-specified timeout, generating syntactically diverse yet semantically equivalent code variants.

\vspace{-0.8ex}
\subsubsection{Mitigating Data Leakage}
\vspace{-0.4ex}

By generating diverse source code variants that differ substantially from publicly available implementations, \toolname mitigates the risk of data leakage from LLM training corpora.
For example, even if an original HeCBench benchmark appeared in an ChatPort's LLM training data, the fuzzed variants introduce novel syntactic patterns, identifier names, and structural arrangements unlikely to have been encountered during training.
This ensures that successful translation reflects genuine understanding rather than memorization, which is particularly important when comparing kernel-level and full-program translation since general-purpose LLMs may have been trained on complete HeCBench programs.

\vspace{-1ex}
\subsection{Runtime Input Fuzzing for Differential Testing}
\vspace{-0.5ex}
\label{subsec:runtime-fuzzing}

The second component of \toolname~focuses on exposing semantic divergences between original and translated implementations through comprehensive runtime testing. While source code fuzzing evaluates translation robustness across diverse input programs, runtime input fuzzing detects behavioral differences that manifest only under specific execution conditions. This component applies uniformly to both translation paths once the translated code compiles successfully.

\vspace{-0.8ex}
\subsubsection{Input Space Exploration}
\vspace{-0.4ex}

HPC kernels typically accept three categories of inputs that affect their execution:
\begin{enumerate}
    \item\textbf{Problem Dimensions.}
Parameters defining array sizes, grid dimensions, block sizes, and iteration counts. These directly impact parallelization strategies, memory access patterns, and workload distribution. For example, the number of threads per block or grid dimensions in CUDA affects how work is distributed across GPU cores.

    \item\textbf{Computational Parameters.}
Numerical values affecting computation such as coefficients, thresholds, timesteps, and configuration constants. These influence algorithmic behavior, numerical stability, and convergence properties. For instance, timestep in molecular dynamics simulations affects both accuracy and stability.

    \item\textbf{Input Data Arrays.}
The actual data arrays processed by the kernel, which may exhibit various statistical properties, value ranges, and patterns. Different data characteristics (e.g., zeros, random values, edge cases, sorted vs. unsorted data) can expose different execution paths and potential bugs in translated code.

\end{enumerate}

\toolname~systematically explores this input space using the following grammar-based fuzzing strategies. Here each of these strategies is customized with input from domain experts to generate \emph{valid} inputs for the application-under-test.

\begin{itemize}
	\item \textbf{Boundary value testing}: Generate edge cases such as minimum and maximum dimension values (e.g., arrays of size 1, very large arrays), zero inputs, extreme numerical values, and boundary conditions that may trigger corner cases in the implementation.

	\item \textbf{Random sampling}: Generate diverse input combinations using pseudo-random sampling with controlled distributions to explore the general input space. This includes varying problem sizes, computational parameters, and data characteristics to exercise different code paths.

	\item \textbf{Mutation-based generation}: Derive new test inputs by perturbing existing inputs, particularly those that revealed interesting behaviors.

	\item \textbf{Coverage-guided prioritization}: Track code coverage to identify under-explored execution paths and prioritize inputs that increase coverage in translated implementation.
\end{itemize}

\vspace{-0.8ex}
\subsubsection{Differential Testing Protocol}
\vspace{-0.4ex}

For each generated runtime input, \toolname~executes both the original and translated implementations and compares their outputs. The protocol addresses several HPC-specific considerations:

\textbf{Cross-Platform Execution.}
The original and translated programs execute on their respective target platforms. For CUDA-to-OpenMP translation, this means executing the original CUDA code on NVIDIA GPUs and the translated OpenMP code on CPUs with OpenMP offloading support.
To enable coverage-guided fuzzing, \toolname uses HeteroBugDetect's~\cite{motwani25} approach that compiles the translated OpenMP program as a static library and creates a \emph{differential driver} program that invokes OpenMP program functions using the static library and CUDA program functions via syatem calls to run the same simulation on two versions using a different set of arguments.
This allows \toolname to measure the code coverage on the translated OpenMP version of the program while executing the simulation on both platforms.

\textbf{Output Comparison with Error Norms.}
Simply comparing outputs with exact equality or basic \texttt{diff} operations generates excessive false positives due to inherent differences in floating-point precision, parallelization strategies, and numerical algorithms across platforms. To address this, \toolname uses error norms commonly used in scientific computing to quantify differences between output arrays:

\begin{itemize}
	\item \textbf{L1 (Manhattan) norm}: $\|x - y\|_1 = \sum_i |x_i - y_i|$ measures the total absolute difference
	\item \textbf{L2 (Euclidean) norm}: $\|x - y\|_2 = \sqrt{\sum_i (x_i - y_i)^2}$ measures root mean square difference
	\item \textbf{Max (Infinity) norm}: $\|x - y\|_\infty = \max_i |x_i - y_i|$ measures maximum pointwise difference
\end{itemize}

A discrepancy is flagged as a correctness failure only when the computed norm exceeds a user-configurable threshold, filtering benign floating-point variations while detecting true semantic divergences.
We set the threshold to 0.0 for computations where the application's own correctness check requires exact equality, and use the application's documented tolerance otherwise.

In practice, \toolname prioritizes the application's built-in verification logic when available, as it reflects domain-appropriate correctness criteria defined by the original developers.
For example, the \emph{lif} application compares spike values against a CPU reference with threshold of $10^{-3}$ tolerance.
When no built-in verification is available, \toolname falls back to one of the three configurable error norms above, with the threshold set by the user based on the application's numerical precision requirements.
In our evaluation, all 18~HeCBench applications include built-in verification, so the built-in checks serve as the primary correctness oracle.

\textbf{Timeout and Resource Monitoring.}
Each execution is subject to timeouts and resource limits to detect infinite loops, deadlocks, or excessive memory consumption that may indicate translation errors. If a translated program takes significantly longer than the original (e.g., more than $10$x), it is recorded as a timeout failure and excluded from correctness analysis.
Full-program translations may introduce performance regressions or resource management issues not present in kernel-only translations, such as inefficient memory transfers or incorrect synchronization that causes excessive waiting.

\textbf{Crash and Exception Handling.}
Runtime errors including segmentation faults, assertion failures, and exceptions are captured and classified as correctness violations. A translation that crashes on any test input is considered incorrect, regardless of whether the original implementation completes successfully.

\textbf{Failure Categorization.}
When a semantic divergence or failure is detected, \toolname~records the specific input that triggered it, the platform configurations used, the error type (crash vs. wrong output), and the magnitude of the divergence (if applicable).
This information enables post-analysis to identify common failure patterns and understand which types of code constructs or input characteristics are most challenging for LLM translation.

For full-program translations, we additionally categorize failures by their location: host-side errors (incorrect memory management, wrong kernel launch parameters) versus device-side errors (incorrect kernel computation). This granular analysis helps identify whether LLMs struggle more with translating computational logic or with translating the orchestration and memory management aspects of HPC programs.

\vspace{-1.80ex}
\section{Evaluation}
\vspace{-0.8ex}
\label{sec:evaluation}

This section describes the dataset, metrics, and implementation and experiment procedure we use to evaluate \toolname, along with the findings of our research questions.

\vspace{-1.80ex}
\subsection{Dataset}
\vspace{-0.5ex}
\label{subsec:dataset}

\toolname is designed to be applicable to any HPC programming model translation task.
In this work, we evaluate it on CUDA-to-OpenMP translation using applications from HeCBench~\cite{Zheming2023}, a large collection of HPC applications written in CUDA, HIP, SYCL, and OpenMP offloading.

We selected a subset of the 47~HeCBench applications used in ChatPORT's evaluation~\cite{ChatPORT_OMP} that satisfy three criteria: (1)~the program executes within one minute on our evaluation platform, enabling practical grammar-based input fuzzing within a 30-minute per-seed budget, (2)~the program includes a reference CPU implementation against which to conduct differential testing, and (3)~correctness results can be computed across all tested LLM variants within the available computational budget.
Of the 47 applications, 21~satisfied these criteria. Three of these (\emph{chemv}, \emph{nbody}, and \emph{axhelm}) could not be evaluated because their CUDA kernels are defined across multiple files and ChatPORT accepts a single file as input.
This is a limitation of the translator under evaluation, not of \toolname.
For two of the remaining applications (\emph{permute} and \emph{chi2}), the source code metamorphic fuzzing generated hundreds of code variants, making it infeasible to compute correctness results across all three LLMs using all code variants within the available computational budget.
The final \toolname evaluation set therefore comprises 16~applications spanning 7~domains including scientific computing, geospatial computing, programming language features, computer vision, deep learning, machine learning primitives, and neuromorphic computing, as shown in Table~\ref{hec_tab02}.

\begin{table}[!t]
	\centering
    \caption{HPC Scientific Applications in the HeCBench benchmark used for \toolname evaluation}
    \label{hec_tab02}
\resizebox{\textwidth}{!}{
	\begin{tabular}{@{}llp{6cm}cc@{}}
		\toprule
		 \textbf{Domain} & \textbf{Application}  & \textbf{Description} & \textbf{Kernel Size(s) (LoC)} & \textbf{Program Size (LoC)} \\
		\midrule
        \multirow{2}{*}{Computational Physics}
        & ace &  Initialization of boundary conditions and swap of grids in the phase-field simulation of dendritic solidification & 23, 8 & 652\\
        & burger & A fundamental partial differential equation used in applied mathematics and physics & 7 & 231 \\
        & ising & Monte-Carlo simulations of 2D Ising model & 22 & 238 \\
        \midrule
        Geospatial Computing
        & aidw & Optimized inverse distance weighting interpolation used in Geographic Information System & 22 & 257 \\
        \midrule
        Programming Language Feature & assert & Evaluate the performance impact of assertion checks in a GPU program & 7 & 80 \\
        \midrule
        \multirow{2}{*}{Computer Vision}
        & background-subtract &  An important pre-processing step in image processing applications such as object tracking & 8, 4, 7, 8 & 180 \\
        & p4 & Post-processing for a 3D object detection network (typical of LiDAR / BEV detectors like PointPillars) & 62 & 223 \\
        & winograd & Optimized convolution operation that shows success in accelerating convolution neural networks, such as VGG and ResNet & 52 & 356 \\
        \midrule
        \multirow{4}{*}{Deep Learning}
        & attention & Scaled dot-product attention for a single query without scaling & 4, 8, 9 & 278 \\
        & channelSum & Per-channel sum of values in a multi-dimensional tensor in deep learning & 17, 16 & 251 \\
        & dense-embedding & Addition of low-dimensional vectors of floating-point numbers that represents a high-dimensional entity in a continuous vector space & 12 & 159 \\
        & flip & A primitive that reverses the order of a tensor along given axis & 20 & 194\\
        & glu & Activation function in artificial neural networks that uses a gating mechanism to control the flow of information & 11 & 148 \\
        \midrule
        \multirow{2}{*}{Machine Learning Primitives}
        & cross &  A primitive that performs cross-product operation of two tensors & 19 & 167 \\
        & nlll & The negative log likelihood loss reduction used in machine learning models for classification tasks & 29 & 216 \\
        \midrule
        Neuromorphic Computing & lif &  A leaky integrate-and-fire neuron model widely used to describe the electrical activity of a biological neuron & 29 & 202 \\
       \bottomrule
        \end{tabular}
    }
\end{table}
\vspace{1ex}

For full-program translation, we evaluate all 47~HeCBench applications from the ChatPORT evaluation set, including the three examples excluded from kernel-level evaluation due to the multi-file limitation.
Since full-program translation does not require kernel extraction, these examples can be included.
We used 5~applications (\emph{Adam}, \emph{Concat}, \emph{Goulash}, \emph{Mrc}, and \emph{Overlay}) for prompt engineering.
To maintain consistency with ChatPORT's evaluation set and enable direct comparison, we evaluated full-program translation on all 47~applications, including the 5~~prompt-engineering applications.
Since prompt engineering only influenced the instruction template and not model weights, the risk of contamination is minimal.

\vspace{-1.80ex}
\subsection{LLM Selection}
\vspace{-0.5ex}
\label{subsec:llms}

We evaluate \toolname using three ChatPORT variants~\cite{ChatPORT_OMP}: $ChatPORT_{\text{CL\_13B}}$ (fine-tuned version of CodeLlama 13B), $ChatPORT_{\text{SCB\_7B}}$ (fine-tuned version of StarCoderBase 7B), and $ChatPORT_{\text{HPC\_C\_6.7B}}$ (fine-tuned version of HPC\_Coder 6.7B).
These were selected from ten ChatPORT variants as the top performers when considering both base model and fine-tuned variant correctness.
$ChatPORT_{\text{CL\_13B}}$ achieved the highest fine-tuned correctness at 72\%, while $ChatPORT_{\text{SCB\_7B}}$ and $ChatPORT_{\text{HPC\_C\_6.7B}}$ each achieved 67\%.
Although $ChatPORT_{\text{SCB\_15B}}$ also reached 72\% after fine-tuning, its base model produced no correct translations (0\%), making $ChatPORT_{\text{SCB\_7B}}$ the stronger overall candidate within the StarCoder family.
All three selected models demonstrated non-trivial baseline capability (9--46\%) in addition to strong fine-tuned performance.

\vspace{-1ex}
\subsection{Metrics}
\vspace{-0.5ex}
\label{subsec:metrics}

We evaluate translation quality using two complementary metrics:

\begin{enumerate}
	\item \textbf{Compilability.}
	The fraction of translated code variants that successfully compile using the target platform's compiler. Formally, given $N$ code variants and their corresponding translations, compilability is defined as:
	\[
	\text{Compilability} = \frac{\text{\# of translations that compile}}{N}
	\]
	This metric measures the LLM's ability to generate syntactically valid code, but it is insufficient alone for evaluating semantic correctness.

	\item \textbf{Correctness.}
	The fraction of translated code variants that compile successfully and produce identical outputs to the original implementation across all fuzzed runtime inputs. Formally, for each compiled translation $T_i$, let $C_i$ denote the set of fuzzed inputs on which $T_i$ produces outputs matching the original implementation. Correctness is defined as:
	\[
	\text{Correctness} = \frac{\sum_{i=1}^{M} \mathbb{1}[|C_i| = |F|]}{N}
	\]
	where $N$ is the number of variants translated, $F$ is the set of all fuzzed runtime inputs, and $\mathbb{1}[\cdot]$ is the indicator function. A translation is considered correct only if it matches the original implementation on \emph{all} fuzzed inputs within the specified error tolerance.

\end{enumerate}

These two metrics provide complementary perspectives: \emph{compilability} assesses syntactic validity and basic platform compatibility, while \emph{correctness} evaluates true semantic preservation. The gap between compilability and correctness reveals the extent to which syntactically valid translations harbor semantic errors---a key insight that motivates our shift from accuracy-focused to correctness-focused evaluation of HPC code translators.
In addition to these quantitative metrics, we perform qualitative error analysis where we analyze compiler error logs and runtime output discrepancies to categorize failure types, characterize their root causes, and identify patterns across LLMs and translation granularities, reporting results as structured taxonomies with representative instances.

\vspace{-1ex}
\subsection{Implementation and Experiment Procedure}
\vspace{-0.5ex}
\label{subsec:implementation}

All experiments were executed on the Oregon State University High-Performance Computing (HPC) cluster using compute nodes from the DGX-2 partition, each equipped with two Intel Xeon Platinum 8168 processors (48 CPU cores total at 2.70 GHz), 16 NVIDIA Tesla V100 GPUs with 32 GB HBM2 memory per GPU (512 GB total GPU memory per node), and 1.5 TB of system memory.
Each node also provided 28 TB of local NVMe storage and high-speed interconnects through 100 Gb Ethernet and Mellanox EDR InfiniBand, with GPUs connected via NVSwitch to enable high-bandwidth GPU-to-GPU communication. This hardware configuration provided a large-scale heterogeneous computing environment suitable for computationally intensive experiments involving parallel CPU and GPU execution.

For source code metamorphic fuzzing, \toolname runs 5~independent fuzzing campaigns per application, each with a 5-minute budget, using the 15~mutation operators described in Section~\ref{subsec:source-fuzzing}, taking approximately 25~minutes per application and 7.5~hours total across all 16~applications. For differential testing using grammar-based input fuzzing, \toolname runs 5 fuzzing campaigns per variant with a 30-minute budget each, taking approximately 2.5~hours per variant. In total, we tested 1,583~unique code variants across all three LLMs, amounting to approximately 3,958~CPU-hours of correctness testing.
Experiments were run in parallel across LLMs on the DGX-2 cluster described above. Output comparison uses the error thresholds documented in each HeCBench application.

\vspace{-1ex}
\subsection{Results}
\vspace{-0.5ex}
\label{subsec:results}

This section presents our findings in terms of the research questions we ask.

\vspace{-0.8ex}
\subsubsection{RQ1: Syntactic vs. Semantic Correctness Gap}
\vspace{-0.4ex}

Table~\ref{tab:rq1_syntactic_semantic} shows the compilability and correctness for kernel-level CUDA-to-OpenMP translations of the 16~HeCBench applications using three ChatPORT variants.
Compilability results are averaged across five metamorphic fuzzing campaigns, while correctness results are computed using source code variants generated from a single representative seed and evaluated using five differential-testing campaigns with runtime inputs; using a single seed for correctness analysis is consistent with ChatPORT's evaluation methodology~\cite{ChatPORT_OMP} and enables direct comparison.

As shown in Table~\ref{tab:rq1_syntactic_semantic}, the compilation-correctness gap is stark and application-dependent.
Several applications achieve near-perfect correctness: \emph{dense-embedding} ($ChatPORT_{\text{CL\_13B}}$: $186/186=1.00$, $ChatPORT_{\text{HPC\_C\_6.7B}}$: $183/186=0.98$), \emph{cross} ($ChatPORT_{\text{CL\_13B}}$: $61/61=1.00$, $ChatPORT_{\text{HPC\_C\_6.7B}}$: $56/61=0.92$), and \emph{glu} ($ChatPORT_{\text{HPC\_C\_6.7B}}$: $178/190=0.94$) demonstrate that high semantic correctness is achievable for some application types.
However, other applications with equally high compilability achieve zero correctness: \emph{lif}~($ChatPORT_{\text{CL\_13B}}$: $393/393=1.00$ compilability, $0/393=0.00$ correctness), \emph{winograd} ($ChatPORT_{\text{CL\_13B}}$: $219/234=0.94$ compilability, $0/252=0.00$ correctness), \emph{flip} ($ChatPORT_{\text{CL\_13B}}$: $226/228=0.99$ compilability, $0/221=0.00$ correctness), and \emph{nlll} ($ChatPORT_{\text{CL\_13B}}$: $136/216=0.63$ compilability, $0/216=0.00$ correctness).
This bimodal distribution confirms that compilation success provides no signal about semantic correctness. A program that compiles and passes developer tests may be perfectly correct or completely wrong, and only differential testing with diverse inputs can distinguish between them.

Testing \emph{burger}, \emph{nlll}, and \emph{lif} applications revealed additional edge cases where the developer-written CUDA program itself failed on \toolname-generated inputs while the LLM-translated OpenMP versions passed, suggesting that the CUDA variants had latent bugs that \toolname exposed independently of the translation.

$ChatPORT_{\text{SCB\_7B}}$ achieves zero compilability on 14 of 16~applications, with meaningful results only for \emph{assert}~(4\%~compilability, 4\%~correctness) and \emph{glu}~(22\%~compilability, 18\%~correctness), making correctness largely unmeasurable for this LLM variant.

\begin{findingbox}
 \textbf{Key Finding:} Neither compilation success nor developer-written tests from static benchmarks can reliably ensure semantic correctness. The results reveal a bimodal pattern: some applications achieve near-perfect correctness (up to 1.00) while others with equally high compilability achieve zero correctness, confirming that shallow evaluation proxies create false confidence in translation quality for scientific HPC applications.~\textbf{(RQ1)}
\end{findingbox}

\begin{table}[!t]
\scriptsize
\caption{Relationship between syntactic and semantic success in LLM-based kernel translation of HPC scientific applications. \textbf{application}: application name; \textbf{Compilability}: avg fraction of OpenMP programs that compile successfully; \textbf{Correctness}: fraction of OpenMP programs that match the outputs on all the tests when compared against the CUDA variants for a single seed(10292).}
\label{tab:rq1_syntactic_semantic}
\begin{tabular}{llrr}
\toprule
\textbf{application} & \textbf{LLM} & \textbf{Compilability} & \textbf{Correctness} \\
\toprule
\multirow{3}{*}{ace} &  $ChatPORT_{CL\_13B}$ & $0/307=0.00$ & N/A  \\
&  $ChatPORT_{SCB\_7B}$ & $0/307=0.00$ & N/A \\
&  $ChatPORT_{HPC\_C\_6.7B}$ & $305/307=0.99$ & $0/308=0.00$ \\
\hline
\multirow{3}{*}{burger} &  $ChatPORT_{CL\_13B}$ & $0/228=0.00$ & N/A  \\
&  $ChatPORT_{SCB\_7B}$ & $0/228=0.00$ & N/A \\
&  $ChatPORT_{HPC\_C\_6.7B}$ & $223/228=0.98$ & $0/208=0.00^{*}$ \\
\hline
\multirow{3}{*}{ising} &  $ChatPORT_{CL\_13B}$ & $191/199=0.96$ & $0/192=0.00$  \\
&  $ChatPORT_{SCB\_7B}$ & $0/199=0.00$ & N/A \\
&  $ChatPORT_{HPC\_C\_6.7B}$ & $0/199=0.00$ & N/A \\
\hline
\multirow{3}{*}{aidw} &  $ChatPORT_{CL\_13B}$ & $102/103=0.99$ & $43/92=0.47$  \\
&  $ChatPORT_{SCB\_7B}$ & $0/103=0.00$ & N/A \\
&  $ChatPORT_{HPC\_C\_6.7B}$ & $102/103=0.99$ & $92/92=1.00$ \\
\hline
\multirow{3}{*}{assert} &  $ChatPORT_{CL\_13B}$ & $0/174=0.00$ & N/A  \\
&  $ChatPORT_{SCB\_7B}$ & $7/174=0.04$ & $6/167=0.04^{*}$ \\
&  $ChatPORT_{HPC\_C\_6.7B}$ & $72/174=0.41$ & $125/167=0.75$ \\
\hline
\multirow{3}{*}{background-subtract} &  $ChatPORT_{CL\_13B}$ & $158/163=0.97$ & $141/155=0.91$  \\
&  $ChatPORT_{SCB\_7B}$ & $0/163=0.00$ & N/A \\
&  $ChatPORT_{HPC\_C\_6.7B}$ & $155/163=0.95$ & $135/155=0.87$ \\
\hline
\multirow{3}{*}{p4} &  $ChatPORT_{CL\_13B}$ & $0/36=0.00$ & N/A  \\
&  $ChatPORT_{SCB\_7B}$ & $0/36=0.00$ & N/A \\
&  $ChatPORT_{HPC\_C\_6.7B}$ & $0/36=0.00$ & N/A \\
\hline
\multirow{3}{*}{winograd} &  $ChatPORT_{CL\_13B}$ & $219/234=0.94$ & $0/252=0.00$  \\
&  $ChatPORT_{SCB\_7B}$ & $0/234=0.00$ & N/A \\
&  $ChatPORT_{HPC\_C\_6.7B}$ & $163/234=0.70$ & $0/252=0.00$ \\
\hline
\multirow{3}{*}{attention} &  $ChatPORT_{CL\_13B}$ & $8/69=0.12$ & $31/65=0.48$  \\
&  $ChatPORT_{SCB\_7B}$ & $0/69=0.00$ & N/A \\
&  $ChatPORT_{HPC\_C\_6.7B}$ & $45/69=0.65$ & $12/63=0.19$ \\
\hline
\multirow{3}{*}{channelSum} &  $ChatPORT_{CL\_13B}$ & $0/95=0.00$ & N/A  \\
&  $ChatPORT_{SCB\_7B}$ & $0/95=0.00$ & N/A \\
&  $ChatPORT_{HPC\_C\_6.7B}$ & $0/95=0.00$ & N/A \\
\hline
\multirow{3}{*}{dense-embedding} &  $ChatPORT_{CL\_13B}$ & $184/188=0.98$ & $186/186=1.00$  \\
&  $ChatPORT_{SCB\_7B}$ & $0/188=0.00$ & N/A \\
&  $ChatPORT_{HPC\_C\_6.7B}$ & $186/188=0.99$ & $183/186=0.98$ \\
\hline
\multirow{3}{*}{flip} &  $ChatPORT_{CL\_13B}$ & $226/228=0.99$ & $0/221=0.00^{*}$  \\
&  $ChatPORT_{SCB\_7B}$ & $0/228=0.00$ & N/A \\
&  $ChatPORT_{HPC\_C\_6.7B}$ & $24/228=0.11$ & $0/221=0.00^{*}$ \\
\hline
\multirow{3}{*}{glu} &  $ChatPORT_{CL\_13B}$ & $179/200=0.90$ & $175/190=0.92$  \\
&  $ChatPORT_{SCB\_7B}$ & $44/200=0.22$ & $35/190=0.18$ \\
&  $ChatPORT_{HPC\_C\_6.7B}$ & $186/200=0.93$ & $178/190=0.94$ \\
\hline
\multirow{3}{*}{cross} &  $ChatPORT_{CL\_13B}$ & $67/67=1.00$ & $61/61=1.00$  \\
&  $ChatPORT_{SCB\_7B}$ & $0/67=0.00$ & N/A \\
&  $ChatPORT_{HPC\_C\_6.7B}$ & $64/67=0.96$ & $56/61=0.92$ \\
\hline
\multirow{3}{*}{nlll} &  $ChatPORT_{CL\_13B}$ & $136/216=0.63$ & $0/216=0.00^{*}$  \\
&  $ChatPORT_{SCB\_7B}$ & $0/216=0.00$ & N/A \\
&  $ChatPORT_{HPC\_C\_6.7B}$ & $82/216=0.38$ & $0/216=0.00^{*}$ \\
\hline
\multirow{3}{*}{lif} &  $ChatPORT_{CL\_13B}$ & $393/393=1.00$ & $0/393=0.00^{*}$  \\
&  $ChatPORT_{SCB\_7B}$ & $0/393=0.00$ & N/A \\
&  $ChatPORT_{HPC\_C\_6.7B}$ & $317/393=0.81$ & $0/393=0.00^{*}$ \\
\hline
\multicolumn{4}{l}{*: \toolname-generated tests inputs found bugs in developer-written CUDA program.}
\end{tabular}
\vspace{-2ex}
\end{table}
\vspace{1ex}

\vspace{-0.8ex}
\subsubsection{RQ2: Taxonomy of Syntactic Errors}
\vspace{-0.4ex}
To identify the types of compilation failures introduced by LLMs during CUDA-to-OpenMP translation, we analyzed compiler errors produced when compiling the OpenMP translations of \toolname-fuzzed CUDA kernels.

Table~\ref{tab:error_taxonomy_kernel_translation} summarizes the resulting taxonomy obtained by analyzing the compilation errors produced by the three ChatPORT variants during \emph{kernel} translations of \toolname-fuzzed kernels of the 16~HeCBench applications.
We identified nine compilation error categories, including \textit{Incorrect Loop Construct}, \textit{Unsupported Construct Usage}, \textit{Function Signature Mismatch}, \textit{Undeclared Identifier}, \textit{Incorrect API/Runtime Name}, \textit{Syntax Error}, \textit{Invalid Control Flow}, \textit{Invalid Construct Usage}, and \textit{Type Mismatch}. These errors primarily arise from the incorrect translation of CUDA kernels into OpenMP work-sharing constructs and are directly reflected in compiler diagnostics.

Several categories identified in our study overlap with error types reported in prior work on LLM-based code translation\cite{Pan24}.
In particular, \textit{Function Signature Mismatch}, \textit{Undeclared Identifier}, \textit{Incorrect API/Runtime Name}, \textit{Syntax Error}, \textit{Type Mismatch}, and \textit{Invalid Control Flow} correspond to common translation failures previously observed across conventional programming languages.
These errors reflect general weaknesses of LLMs, including omitted declarations, inconsistent identifier generation, incorrect API substitution, malformed syntax, and type-related reasoning failures.

However, our analysis also reveals error categories that are largely specific to CUDA-to-OpenMP translation.
In particular, \textit{Incorrect Loop Construct}, \textit{Unsupported Construct Usage}, and \textit{Invalid Construct Usage} stem from the semantic gap between CUDA's hierarchical execution model and OpenMP's directive-based programming model.
These errors include invalid loop-collapse structures, illegal nesting of OpenMP regions, misuse of reduction and atomic clauses, and unsupported combinations of OpenMP directives and clauses.
Such errors have not been reported in prior code-translation studies~\cite{oxidizer2025}
as our identified error types arise from translating between heterogeneous parallel programming models rather than between conventional sequential programming languages (e.g, Go-to-Rust).
\vspace{-2ex}

\begin{table}[!t]
\scriptsize
\caption{Compilation errors produced by three top-performing ChatPORT~\cite{ChatPORT_OMP} LLMs (CodeLlama\_13B~(CL), StarCoderBase\_7B~(SC), HPC\_Coder\_6.7B~(HC)) for CUDA-to-OpenMP kernel translations when evaluated on \toolname-fuzzed kernels.}
\label{tab:error_taxonomy_kernel_translation}

\begin{tabular}{p{2.5cm}p{3cm}p{4cm}p{3.4cm}}
\toprule
\textbf{Error Type} & \textbf{Description} & \textbf{Error Instance} & \textbf{Reasoning} \\
\midrule

Incorrect loop construct &
Loop structure violates OpenMP loop requirements. &
\texttt{ace(CL)}, \texttt{ising(CL)}: insufficient nested loops for \texttt{collapse(N)}; \texttt{glu(CL)}, \texttt{channelSum(HC)}: non-\texttt{for} loop after OpenMP directive; \texttt{assert(CL)}, \texttt{attention(CL)}: invalid loop condition. &
LLM generated fewer nested loops than required, used unsupported loop forms, or copied CUDA loop logic incompatible with OpenMP. \\
\midrule

Unsupported construct usage &
OpenMP constructs or clauses used in invalid contexts. &
\texttt{ace(CL)}, \texttt{channelSum(CL)}, \texttt{flip(SC)}: illegal region nesting; \texttt{channelSum(CL)}: invalid \texttt{num\_teams} clause; \texttt{dense-embedding(CL)}: invalid \texttt{grainsize} clause. &
LLM applied directives or clauses in contexts where they are not permitted by OpenMP. \\
\midrule

Function signature mismatch &
Function calls do not match available definitions. &
\texttt{channelSum(CL)}, \texttt{ace(HC)}, \texttt{flip(SC)}: no matching function for translated kernel calls; \texttt{aidw(SC)}: unresolved overloaded function. &
LLM translated function invocations without generating compatible definitions or correct argument signatures. \\
\midrule

Undeclared identifier &
Variable, macro, constant, or function referenced without declaration. &
\texttt{attention(CL)}, \texttt{glu(HC)}, \texttt{dense(SC)}: fabricated identifiers; \texttt{ace(CL)}, \texttt{burger(HC)}, \texttt{lif(SC)}: missing variable declarations; \texttt{ising(CL)}, \texttt{winograd(HC)}: misspelled identifiers. &
LLM omitted declarations, altered identifiers during translation, or generated placeholder names absent from the original code. \\
\midrule

Incorrect API/runtime name &
CUDA-specific APIs, types, or runtime variables retained in OpenMP code. &
\texttt{assert(CL)}, \texttt{channelSum(HC)}, \texttt{ace(SC)}: use of \texttt{gridDim}/\texttt{blockIdx}; \texttt{ising(CL)}, \texttt{glu(SC)}: unsupported CUDA types (\texttt{int3}, \texttt{float4}). &
CUDA constructs were copied directly instead of being translated into valid OpenMP/C++ equivalents. \\
\midrule

Syntax error &
Generated code is not syntactically valid C++. &
\texttt{nlll(CL)}, \texttt{cross(HC)}, \texttt{lif(SC)}: mismatched braces and missing expressions; \texttt{aidw(SC)}, \texttt{assert(SC)}: invalid declarations; \texttt{flip(HC)}: malformed type declarations. &
Incomplete or malformed code generation produced unparsable source code. \\
\midrule

Invalid control flow &
Control-flow statements violate OpenMP execution rules. &
\texttt{background-subtract(CL)}, \texttt{glu(HC)}, \texttt{ising(SC)}: \texttt{return} inside OpenMP region; \texttt{nlll(HC)}, \texttt{dense(SC)}: \texttt{break} inside OpenMP loop. &
LLM generated control flow that exits parallel regions in ways prohibited by OpenMP. \\
\midrule

Invalid construct usage &
Incorrect use of OpenMP clauses or synchronization constructs. &
\texttt{glu(CL)}: invalid \texttt{atomic} updates; \texttt{nlll(CL)}, \texttt{glu(SC)}: invalid reduction variables; \texttt{channelSum(HC)}: invalid \texttt{num\_threads} argument. &
LLM violated OpenMP semantic requirements for reductions, atomics, data movement, or clause placement. \\
\midrule

Type mismatch &
Operations performed on incompatible types. &
\texttt{glu(CL)}, \texttt{flip(SC)}: incompatible operands; \texttt{winograd(CL)}: indexing function pointer; \texttt{lif(CL)}: pointer-scalar assignment; \texttt{ace(SC)}: string used as integral value. &
Translation introduced incompatible types, incorrect casts, or naming conflicts that changed expression semantics. \\
\bottomrule
\end{tabular}

\end{table}
\vspace{1ex}

\begin{findingbox}
 \textbf{Key Finding:} Kernel-level LLM translations produce nine compile-time error categories: \emph{Incorrect Loop Construct, Unsupported Construct Usage, Function Signature Mismatch, Undeclared Identifier, Incorrect API/Runtime Name, Syntax Error, Invalid Control Flow, Invalid Construct Usage,} and \emph{Type Mismatch.}
 While several of these align with general LLM translation failures documented in prior work, \emph{Incorrect Loop Construct, Unsupported Construct Usage, and Invalid Construct Usage} are unique to the semantic gap between CUDA's hierarchical execution model and OpenMP's directive-based programming model. \textbf{(RQ2)}
\end{findingbox}

\vspace{-0.8ex}
\subsubsection{RQ3: Taxonomy of Semantic Errors}
\vspace{-0.4ex}

\begin{table}[!t]
\centering
\caption{Runtime Errors produced by three LLMs (CodeLlama\_13B~(CL), StarCoderBase\_7B~(SC), HPC\_Coder\_6.7B~(HC)) using the kernel-level fine-tuned variants from ChatPORT~\cite{ChatPORT_OMP} for CUDA-to-OpenMP kernel translations using \toolname-fuzzed kernels.}
\label{tab:error_taxonomy_kernel_translation_runtime}
\scriptsize
\begin{tabular}{p{3cm}p{6cm}p{5cm}}
\toprule
\textbf{Category} & \textbf{Description} & \textbf{Affected Benchmarks} \\
\midrule

Intermediate Variable Elimination &
A computed intermediate variable is dropped and a raw input substituted in its place, silently altering formula semantics. The program compiles and runs but produces wrong numerical output. &
(aidw): \texttt{R\_S0} (normalized observation ratio) is dropped; raw \texttt{r\_obs} is substituted directly. All threshold comparisons and the cosine argument operate at the wrong scale, corrupting \texttt{u\_R}, \texttt{alpha}, and every output point. \\
\hline

Execution Model Assumption Transfer &
An implicit CUDA execution guarantee --- exact thread count per block, block-to-batch isolation, or nested parallelism hierarchy --- is carried into OpenMP without the guards or restructuring needed to make it hold at runtime. &
(aidw): \texttt{tid} computed as \texttt{team * BLOCK\_SIZE + lid}, but the OMP runtime silently reduces team thread count below \texttt{BLOCK\_SIZE}. Output indices are skipped and shared-memory tiles are partially uninitialized.

(dense embedding): \texttt{target teams distribute + parallel for} does not guarantee \texttt{batch\_idx} isolation per team. The runtime may expose a wrong \texttt{batch\_idx} to inner threads, producing incorrect \texttt{dense\_elem} lookups. \\
\hline

Loop Bound Error &
\texttt{threadIdx.x} is replaced by an explicit sequential loop, but the loop bound is set to the hardware block-size parameter rather than the data dimension, leaving output elements beyond the block size permanently unwritten. &
(dense embed, dense\_esuhm): \texttt{threadIdx.x} serialized as \texttt{for (tid=0; tid<numThreads; tid++)}. With \texttt{numThreads=128} and \texttt{embedding\_dim=768}, columns 128--767 (640 of 768) are never written. \\

\hline
Missing Statement Fault & A mandatory in-place state mutation is silently dropped during translation. The affected variable is used immediately in downstream computation and written back to global memory, so the stale value corrupts both the current output and all subsequent time steps that read the same location. &
(lif): \texttt{ref\_time -= dt} is omitted in the OpenMP translation. The refractory multiplier \texttt{tmp\_val529} and the spike-triggered \texttt{ref\_time} reset both use the un-decremented value. With the error accumulating across steps, \texttt{spikes} diverges from the CUDA output by step~1 for \texttt{neurons\_per\_item=1000}, \texttt{num\_items=32}, \texttt{num\_steps=1000}. \\

\hline
Shared Memory Scope Mistranslation &
CUDA \texttt{\_\_shared\_\_} arrays are translated into variables declared at the \texttt{teams} scope rather than inside the \texttt{parallel} region with explicit shared visibility. Inter-thread access to these variables is implementation-defined in OpenMP, causing the thread-0 reduction to read uninitialized or privatized values and producing wrong output. &
(nlll): \texttt{sm\_inputs} and \texttt{acc\_weight} declared in the \texttt{target teams} region outside \texttt{parallel}. The per-thread accumulation writes are not guaranteed visible to thread~0 at reduction time, corrupting both \texttt{output} and \texttt{total\_weight} for input \texttt{(8192, 1000, 10)}. \\

\hline
Multi-dimensional Index Flattening &
A multi-dimensional thread index space is incorrectly translated to a single linear index expression applied uniformly across all dimensions. Independent spatial axes that should resolve to different values become identical, causing threads to compute wrong coordinates and write to wrong or duplicate output locations. &
(winograd): \texttt{tile\_i} and \texttt{tile\_j} are both computed as \texttt{omp\_get\_team\_num() * omp\_get\_num\_threads() + omp\_get\_thread\_num()}, collapsing the 2D CUDA grid (\texttt{blockIdx.x/y}, \texttt{threadIdx.x/y}) onto a single 1D index. Every thread processes a diagonal point where \texttt{tile\_i == tile\_j}, leaving all off-diagonal output elements unwritten for input \texttt{(32, 8, 8)}. \\

\bottomrule

\end{tabular}
\end{table}
\vspace{1ex}

Table~\ref{tab:error_taxonomy_kernel_translation_runtime} presents the taxonomy of semantic errors identified through \toolname's differential testing of kernel-level translations across the 16~HeCBench applications.
We identified six categories of semantic errors that survive compilation and produce incorrect scientific results only under specific input conditions.

The first category, \textit{Intermediate Variable Elimination}, occurs when a computed intermediate variable is silently dropped during translation and a raw input substituted in its place, altering formula semantics without triggering any compilation error.
The second category, \textit{Execution Model Assumption Transfer}, arises when implicit CUDA execution guarantees, such as exact thread count per block or block-to-batch isolation, are carried into OpenMP without the guards or restructuring needed to hold at runtime.
The third category, \textit{Loop Bound Error}, occurs when \texttt{threadIdx.x} is serialized into an explicit loop but the loop bound is set to the hardware block-size parameter rather than the data dimension, leaving output elements beyond the block size permanently unwritten.
The fourth category, \textit{Missing Statement Fault}, covers cases where a mandatory in-place state mutation is silently omitted during translation; the affected variable is used immediately in downstream computation and written back to global memory, so the stale value corrupts both the current output and all subsequent timesteps that read the same location.
For example, as described in Section~\ref{sec:motivation-example}, the LIF bug is an instance of this category where \texttt{$ref\_time -= dt$} was omitted from the OpenMP translation, and \toolname generated multiple diverse inputs to expose this fault.
The fifth category, \textit{Shared Memory Scope Mistranslation}, occurs when CUDA \texttt{\_\_shared\_\_} arrays are translated into variables declared at the teams scope rather than inside the parallel region with explicit shared visibility, causing inter-thread access to be implementation-defined in OpenMP and producing wrong output.
The sixth category, \textit{Multi-dimensional Index Flattening}, occurs when a multi-dimensional thread index space is incorrectly translated to a single linear index expression applied uniformly across all dimensions, collapsing independent spatial axes onto a single diagonal and leaving all off-diagonal output elements unwritten.

All six categories share a common characteristic: they are input-dependent. Each error manifests only under specific combinations of array sizes, computational parameters, or simulation durations, and none were exposed by the developer-written seed inputs provided with the HeCBench applications.
Notably, the semantic errors in \emph{dense-embedding} are LLM-specific: $ChatPORT_{\text{CL\_13B}}$ achieves 1.00~correctness while other ChatPORT variants exhibited \emph{Loop Bound Error} and \emph{Execution Model Assumption Transfer}, confirming that different LLMs handle the same semantic translation challenges differently.
These findings confirm that behavioral testing with diverse, \toolname-generated inputs is necessary to detect semantic errors in LLM-translated HPC applications.

Comparing the semantic error taxonomy with the compile-time error taxonomy (Table~\ref{tab:error_taxonomy_kernel_translation}) reveals a qualitative shift in error character: the nine compile-time categories are syntactic and structural, immediately surfaced by the compiler, whereas the six semantic categories are behaviorally silent at compile time and manifest only under specific runtime conditions, making them substantially more dangerous in practice.
ChatPORT~\cite{ChatPORT_OMP} reports four compile-time error categories, all of which correspond to categories in our RQ2 taxonomy, but does not report any semantic errors since its evaluation relies on compilation success and fixed developer-written seed tests.

Comparing our semantic error taxonomy against the seven heterogeneous bug categories identified by HeteroBugDetect~\cite{motwani25} for platform-specific divergences in developer-written CUDA/OpenMP code: \emph{incorrect host-device synchronization, missing host-device synchronization, accessing device memory from host, missed data copying, incorrect data copying, use of stale data,} and \emph{concurrent modification of shared variables}.
Our \textit{Execution Model Assumption Transfer} category shares conceptual overlap with HeteroBugDetect's missing-synchronization and stale-data categories, as both arise when an implicit guarantee about execution order or shared state fails to hold once code crosses a platform boundary.
However, our remaining two categories, \textit{Intermediate Variable Elimination} and \textit{Loop Bound Error}, have no clear analog in HeteroBugDetect's taxonomy. These errors arise from the LLM's translation process itself, specifically dropped computational logic and mis-set loop bounds during thread-indexing serialization, rather than from manual host-device memory management mistakes.
This suggests that LLM-translated code exhibits a partially distinct error profile from developer-written heterogeneous code: while both share synchronization and state-consistency risks, LLM translation additionally introduces code-generation-level errors that have no counterpart in bugs arising from manual heterogeneous programming.
To our knowledge, the six-category taxonomy presented here is the first characterization of semantic errors in LLM-based CUDA-to-OpenMP translation.
\vspace{-1ex}

\begin{findingbox}
\textbf{Key Finding:} Semantic errors in LLM-translated HPC programs are input-dependent and elude fixed developer-written tests. \toolname's differential testing exposes six categories of semantic errors: \emph{Intermediate Variable Elimination, Execution Model Assumption Transfer, Loop Bound Error, Missing Statement Fault, Shared Memory Scope Mistranslation, and Multi-dimensional Index Flattening.}
These errors share a common pattern: they preserve the program's structural appearance while silently altering its computational behavior, and they manifest only under specific runtime conditions that fixed developer tests do not exercise. \textbf{(RQ3)}
\end{findingbox}

\vspace{-0.8ex}
\subsubsection{RQ4: Translation Granularity vs. LLM Success}
\vspace{-0.4ex}

Table~\ref{tab:rq4_granularity} compares kernel-level and full-program translation outcomes across the three ChatPORT variants on all 47~HeCBench applications from the ChatPORT evaluation set~\cite{ChatPORT_OMP}, including the 5~examples used for prompt engineering.
Kernel-level translation achieves non-trivial correctness rates across all three LLMs, with $ChatPORT_{\text{CL\_13B}}$ achieving the highest correctness at 72\%.

Full-program translation is substantially harder. $ChatPORT_{\text{CL\_13B}}$ fails to compile entirely (0\%), while $ChatPORT_{\text{SCB\_7B}}$ and $ChatPORT_{\text{HPC\_C\_6.7B}}$ achieve only 2\% and 28\% compilability respectively, and at most 15\% correctness. Notably, kernel-level fine-tuning degrades both compilability and correctness for $ChatPORT_{\text{CL\_13B}}$ (0\% vs~base model's~4\% compilability, 0\% vs~2\% correctness) and $ChatPORT_{\text{SCB\_7B}}$ (2\% vs~15\% compilability, 2\% vs~6\% correctness), suggesting over-specialization toward kernel-only output.
$ChatPORT_{\text{HPC\_C\_6.7B}}$ is the only variant where fine-tuning improves both compilability (28\%~vs~23\%) and correctness (15\%~vs~6\%), though absolute correctness remains low.
These results confirm that full-program translation remains an unsolved challenge even for fine-tuned models.

\begin{table}[!t]
\centering
\scriptsize
\caption{The effect of translation granularity on LLM-based CUDA-to-OpenMP translation across 47~HPC applications from HeCBench.
The ``Full-Program'' shows the performance when translating the entire CUDA program, while the ``Kernel-Only'' shows the performance of translating only the CUDA kernels of the applications; the results of ``Kernel-Only'' are borrowed from the ChatPORT~\cite{ChatPORT_OMP}.}
\label{tab:rq4_granularity}
\begin{tabular}{rrrr}
\toprule
\textbf{LLM} & \textbf{Granularity} & \textbf{Compilability} & \textbf{Correctness}  \\
\midrule
\multirow{2}{*}{$ChatPORT_{CL\_13B}$} & Full-Program & $0/47=0.00$ & $0/47=0.00$ \\
& Kernel-Only & N/A & $0.72$ \\
\midrule
\multirow{2}{*}{$CodeLlama\_13B$} & Full-Program & $2/47=0.04$ & $1/47=0.02$ \\
& Kernel-Only & N/A & $0.22$ \\
\midrule
\midrule
\multirow{2}{*}{$ChatPORT_{SCB\_7B}$} & Full-Program & $1/47=0.02$ & $1/47=0.02$ \\
& Kernel-Only & N/A & $0.67$ \\
\midrule
\multirow{2}{*}{$StarCoderBase\_7B$} & Full-Program & $7/47=0.15$ & $3/47=0.06$ \\
& Kernel-Only & N/A & $0.09$ \\
\midrule
\midrule
\multirow{2}{*}{$ChatPORT_{HPC\_C\_6.7B}$} & Full-Program & $13/47=0.28$ & $7/47=0.15$ \\
& Kernel-Only & N/A & $0.67$ \\
\midrule
\multirow{2}{*}{$HPC\_Coder\_6.7B$} & Full-Program & $11/47=0.23$ & $3/47=0.06$ \\
& Kernel-Only & N/A & $0.46$ \\
\bottomrule
\end{tabular}
\end{table}
\vspace{1ex}

\begin{table}[!t]
\scriptsize
\caption{Compilation errors produced in CUDA-to-OpenMP full program translations of original HeCBench applications using three top-performing LLMs (CodeLlama\_13B~(CL), StarCoderBase\_7B~(SC), HPC\_Coder\_6.7B~(HC)) from ChatPORT~\cite{ChatPORT_OMP}.}
\label{tab:error_taxonomy_full_translation}
\resizebox{\textwidth}{!}{
\begin{tabular}{p{3cm} l p{5.4cm} p{4.0cm}}

\toprule
\textbf{Category} & \textbf{Error Type} & \textbf{Description} & \textbf{Error Instances} \\
\midrule

\textbf{Offloading} &
Missing offloading &
Missing OMP pragma offloading, or failure to translate CUDA kernel launches to \texttt{target} directives. &
\textit{gd} (CL); \textit{ising} (CL,SC); \textit{mrc} (CL); \textit{nbody} (CL,SC,HC) \\
\cline{2-4}
&
Incorrect offloading &
Wrong OpenMP constructs (e.g., wrong \texttt{collapse}, incorrect thread limits), or nonsensical thread counts. &
\textit{vol2col} (CL,SC,HC); \textit{iso2dfd} (CL,SC); \textit{lif} (CL,SC,HC); \textit{cmp} (HC) \\[1pt]
\hline
\textbf{Memory Management} &
Missing memory management &
No OMP mapping, missing allocation/deallocation, or missing shared memory. &
\textit{haccmk} (CL,SC); \textit{hotspot3d} (all); \textit{ising} (CL,SC); \textit{iso2dfd} (CL,SC); \textit{lif} (all); \textit{mcpr} (all); \textit{mrc} (all); \textit{nbody} (SC,HC) \\
\cline{2-4}
&
Incorrect memory management &
Incorrect host-device mapping or data movement; incorrect shared memory simulation. &
\textit{winograd} (all); \textit{haccmk} (SC,HC) \\
\cline{2-4}
&
Redundant data allocation &
Redundant memory allocation via OpenMP or extra variables. &
\textit{burger} (CL,SC); \textit{haccmk} (HC) \\[1pt]
\hline
\textbf{Parallelism} &
Incorrect parallel runtime semantics &
Incorrect loop nesting or host-device execution flow. &
\textit{atomicCost} (SC); \textit{chi2} (SC) \\
\cline{2-4}
&
Missing unrolling &
Missing manual loop unrolling or reduction directives. &
\textit{axhelm} (all) \\
\cline{2-4}
&
Incorrect parallel indexing &
Incorrect indexing for OMP threads or teams. &
\textit{ace} (CL,SC); \textit{vol2col} (CL,SC) \\
\cline{2-4}
&
Missing data synchronization &
Missing OMP atomics; incorrect concurrent variable updates. &
\textit{gd}; \textit{AIDW} (HC) \\[1pt]
\hline

\textbf{Code Validity} &
Hallucinated APIs &
Non-existent functions or APIs; device-only calls invoked on host. &
\textit{aidw} (SC); \textit{attention} (SC,HC) \\
\cline{2-4}
&
CUDA code/header retention &
Residual CUDA API calls or headers in translated code. &
\textit{lif}; \textit{ACE} (CL,SC); \textit{particle-diffusion} (SC); \textit{axhelm} (CL,SC) \\
\cline{2-4}
&
Missing macro definitions &
Undefined constants or missing preprocessor macros. &
\textit{burger} (CL) \\
\cline{2-4}
&
Algorithmic errors &
Logic errors changing computational results (e.g., wrong filter or hardcoded dimensions). &
 \textit{flip} (SC) \\
\cline{2-4}
 &
Missing dependencies/headers &
Missing \texttt{\#include} statements for required headers. &
\textit{overlay} (CL); \textit{nll} (CL,SC) \\
\cline{2-4}
&
Type inconsistency &
Wrong data types (e.g., \texttt{int} vs.\ \texttt{float}/\texttt{DATA\_TYPE}). &
\textit{iso2dfd} (HC); \textit{mcpr} (all); \textit{nbody} (all) \\
\cline{2-4}
&
Incorrect function signatures &
Parameter type/count mismatch or incorrect function definition. &
\textit{vol2col} (SC); \textit{assert} (CL) \\
\cline{2-4}
&
Missing variable declarations &
Variables used without declaration; undeclared identifiers. &
\textit{channelShuffle} (SC); \textit{aidw} (SC) \\
\cline{2-4}
&
Syntax errors &
Invalid C/C++ syntax preventing compilation. &
\textit{swish} (SC) \\[1pt]
\cline{2-4}
 &
Incorrect array access &
Wrong indexing schemes or dimensionality mismatches. &
\textit{backprop} (SC,HC) \\
\cline{2-4}
 &
Hardcoded values &
Literals used instead of symbolic constants or parameters. &
\textit{backprop} (SC,HC) \\
\cline{2-4}
 &
Incomplete implementations &
Stub functions or partially translated kernels. &
\textit{fdtd3d} (HC) \\
\cline{2-4}
 &
Missing function implementations &
Functions entirely absent from translated output. &
\textit{chi2} (CL); \textit{page-rank} (CL); \\[1pt]

\hline
\textbf{Performance} &
Incorrect timing/measurement &
Timing present but incorrectly implemented. &
\textit{nbody} (CL,SC); \textit{channelShuffle} (HC) \\
\cline{2-4}
&
Missing timing/measurement &
Timing/measurement mechanism entirely absent. &
\textit{background-subtract} (CL) \\[1pt]

\hline
\textbf{Compatibility} &
Multi-file compatibility errors &
Type or signature mismatches across translation units. &
\textit{nbody} (CL,SC) \\
\cline{2-4}
&
Missing/incorrect verification &
Inadequate result validation or missing error checks. &
\textit{assert} (CL,SC) \\
\cline{2-4}

&
Naming convention inconsistencies &
Function/variable names differ from expected conventions. &
\textit{page-rank} (SC,HC) \\
\bottomrule
\multicolumn{4}{l}{\footnotesize all\,=\,CL,\,SC,\,HC}
\end{tabular}
}
\vspace{-3ex}
\end{table}
\vspace{1ex}

Table~\ref{tab:error_taxonomy_full_translation} presents the 27-category taxonomy of compile-time errors identified across \emph{full-program} translation of all 47~evaluated applications.
As shown, we observed a broader set of compilation and translation errors spanning \emph{offloading, memory management, parallelism, code validity, performance instrumentation,} and \emph{compatibility} concerns.
While many of these categories manifest through the same compiler diagnostics observed in kernel translations, full-program translation introduces additional challenges related to host-device memory management, OpenMP offloading directives, multi-file dependencies, and incomplete program implementations.
For example, categories such as \textit{Missing Offloading}, \textit{Incorrect Memory Management}, and \textit{Missing Function Implementations} do not typically arise in isolated kernel translation because they involve interactions between kernels, host code, and application-level infrastructure.

The qualitative difference between the two granularities is reflected in the error profiles.
Comparing the two translation granularities reveals that kernel-level translation failures are dominated by low-level syntactic compilation errors, whereas full-program translation introduces additional system-level concerns related to offloading, memory management, dependency resolution, and application integration.
These findings suggest that successful kernel translation alone is insufficient for achieving correct end-to-end migration of scientific applications, as full-program translation requires reasoning about both parallel execution semantics and application-level infrastructure.

\begin{findingbox}
    \textbf{Key Finding:} Full-program translation is substantially harder than kernel-level translation. While kernel-level translation achieves up to 72\% correctness, full-program translation fails to compile entirely for $ChatPORT_{\text{CL\_13B}}$, with errors spanning 27~categories across \emph{offloading, memory management, parallelism, code validity, performance,} and \emph{compatibility}.
    Kernel-level fine-tuning degrades both compilability and correctness for full-programs using $ChatPORT_{\text{CL\_13B}}$ (0\% vs~4\% compilability, 0\%~vs~2\% correctness) and $ChatPORT_{\text{SCB\_7B}}$ (2\%~vs~15\% compilability, 2\% vs~6\% correctness), while only $ChatPORT_{\text{HPC\_C\_6.7B}}$ benefits from fine-tuning (28\%~vs~23\% compilability, 15\% vs~6\% correctness). \textbf{(RQ4)}
\end{findingbox}

\vspace{-1.5ex}
\section{Related Work}
\vspace{-0.8ex}
\label{sec:relatedwork}

This section places our contributions in the context of existing research.

\textbf{LLM-Based HPC Code Translation.}
Recent advances in large language models (LLMs) have enabled automated code translation across programming languages and software ecosystems.
Prior work has shown that LLMs can successfully translate code between conventional programming languages, but often introduce errors such as syntax violations, missing declarations, type inconsistencies, incorrect API mappings, and function signature mismatches~\cite{Pan24}.
To improve translation quality at the project level, recent frameworks such as Oxidizer~\cite{oxidizer2025} employ feature mapping and type-compatibility validation to translate entire software projects while mitigating common translation failures.
However, existing studies primarily focus on sequential programming languages and conventional software systems.
c
In the HPC domain, LLMs have recently been explored for translating accelerator programming models, including CUDA-to-OpenMP migration.
Several recent tools use LLMs to automate translation between HPC programming models. CodeRosetta~\cite{coderosetta} uses an encoder-decoder transformer trained on CUDA and SYCL code pairs to translate between the two models. Fortran2CPP~\cite{fortran2cpp} fine-tunes LLMs specifically for Fortran-to-C++ translation, targeting legacy scientific codebases. LASSI~\cite{lassi} and UniPar~\cite{unipar} take broader approaches to parallelism-aware code transformation using LLM-based techniques. ChatPORT~\cite{ChatPORT_OMP, ChatPORT_SYCL}, developed as part of this research program, fine-tunes a range of open-source code LLMs for CUDA-to-OpenMP and CUDA-to-SYCL kernel translation, achieving correctness rates of up to 79\% and 81.7\% respectively.
A common thread across all of these approaches is reliance on compilation success, token-level similarity metrics such as CodeBLEU~\cite{codebleu}, and developer-written tests from static benchmarks for evaluation. \toolname complements these translation tools by providing a correctness-oriented evaluation framework that goes beyond syntactic proxies to assess behavioral equivalence.

Our error taxonomies complement and extend prior work in two ways.
First, several categories in our kernel-level compile-time taxonomy, including \textit{Function Signature Mismatch}, \textit{Undeclared Identifier}, \textit{Incorrect API/Runtime Name}, \textit{Syntax Error}, \textit{Type Mismatch}, and \textit{Invalid Control Flow}, correspond to general LLM translation failures reported by Pan et al.~\cite{Pan24} and addressed by project-scale frameworks such as Oxidizer~\cite{oxidizer2025}.
Second, three categories, \textit{Incorrect Loop Construct}, \textit{Unsupported Construct Usage}, and \textit{Invalid Construct Usage}, are specific to the semantic gap between CUDA's hierarchical execution model and OpenMP's directive-based model and have not been reported in prior studies of sequential language translation.
ChatPORT~\cite{ChatPORT_OMP} reports four compile-time error categories, all of which correspond to categories in our kernel-level taxonomy, but does not characterize semantic errors since its evaluation relies on compilation success and fixed developer-written seed tests.
Our semantic error taxonomy, comprising six categories including \textit{Execution Model Assumption Transfer}, \textit{Shared Memory Scope Mistranslation}, and \textit{Multi-dimensional Index Flattening}, is to our knowledge the first characterization of runtime errors in LLM-based CUDA-to-OpenMP translation.

\textbf{Testing and Verification of HPC Programs.}
The evolution of testing in HPC has progressed from early efforts focused on numerical consistency and basic system correctness toward modern, multi-layered methodologies that address the complexity of heterogeneous, massively parallel architectures.
Foundational work on floating-point determinism, such as FLiT’s cross-platform result-consistency testing~\cite{sawaya2017flit} and ReproBLAS’s efficient, mathematically bounded summation routines~\cite{demmel2016reproblas}, demonstrated the need for rigorous numerical validation in environments where non-associativity and parallel reduction order introduce nondeterminism. These foundational concerns directly motivate \toolname's use of configurable error norms rather than exact equality for output comparison.

As systems scaled out, correctness efforts extended to communication-centric verification frameworks like MUST~\cite{protze2012must} and ISP~\cite{gopalakrishnan2008isp}, which enabled scalable detection of MPI errors such as deadlocks, datatype mismatches, and ordering bugs, while performance regression testing emerged as a vital component of software quality, with methodologies that systematically compare performance profiles across runs to detect regressions induced by hardware, kernel, or compiler changes~\cite{shende2006perf}.
In parallel, research on system noise and jitter showed how operating-system and hardware variability can significantly impact scalability~\cite{ferreira2008noise}, motivating HPC centers to refine acceptance and health-check suites for processors, interconnects, and runtime environments.
\toolname's timeout and resource monitoring mechanism addresses similar concerns in the context of LLM-translated code.

As workflows and software stacks became more complex, the community increasingly emphasized reproducibility as a testing dimension in its own right: tools like ReproZip use provenance to capture and replay complete computational environments~\cite{chirigati2013reprozip}, while best-practice guidelines for computational science advocate infrastructure and processes that make results reproducible and extensible across platforms and time~\cite{stodden2014best}.
Storage- and I/O-layer correctness also gained prominence, with empirical analyses of latent sector errors and data corruption in disks~\cite{bairavasundaram2008corruption} informing the design of more robust parallel filesystems and stress-testing tools.
Collectively, these developments reflect a field that has evolved from isolated correctness checks into a holistic, system-wide testing ecosystem spanning numerical accuracy, communication correctness, performance stability, resilience, storage integrity, and workflow reproducibility.
\toolname builds on this tradition of rigorous HPC correctness verification by applying differential testing to evaluate LLM-translated code rather than developer-written implementations.

Most closely related to \toolname are techniques such as HeteroBugDetect~\cite{motwani25} and HeteroFuzz~\cite{Zhang21}, which detect semantic divergences between implementations of the same program running on different hardware.
\toolname builds on HeteroBugDetect's differential testing methodology by evaluating LLM-translated codes rather than developer-written implementations, and by combining it with metamorphic source code fuzzing to generate diverse translation inputs that mitigate data leakage and expose translation robustness gaps.

\textbf{Metamorphic Testing.}
Metamorphic testing~\cite{chen1998metamorphic} detects bugs by verifying that semantics-preserving input transformations produce consistent outputs.
It has been widely applied to complex systems including search engines, compilers, and web APIs~\cite{segura2016, xiao2022, tao2010}.
In the context of LLM-based systems, metamorphic testing has been used to evaluate robustness of neural machine translation~\cite{Steven25}.
\toolname applies metamorphic testing at the source code level, generating semantically equivalent program variants as translation inputs.
This is distinct from prior applications that apply metamorphic relations at the input-output level of a fixed program.
By testing LLM translation robustness across a large corpus of semantically equivalent program variants, \toolname exposes translation failures that evaluation on single fixed programs cannot detect.

\textbf{Differential Testing and Fuzzing for Code.}
Differential testing was introduced by McKeeman~\cite{McKeeman1998} and has been highly effective for compiler testing.
CSmith~\cite{yang2011} generates random C programs and uses differential testing across compilers to find miscompilation bugs, finding hundreds of bugs in GCC and LLVM.
Equivalence Modulo Inputs (EMI) testing~\cite{Sun2016} generates semantically equivalent program variants by inserting dead code into live regions, exploiting the same principle as our source code fuzzing approach.
Grammar-based fuzzing has been applied broadly to programs, protocols, and file formats~\cite{Aschermann2019}.
\toolname combines these ideas in a novel way: grammar-based source code fuzzing generates diverse translation inputs to test LLM robustness, while runtime input fuzzing combined with differential testing detects semantic divergences in translated programs.
Unlike compiler testing tools that compare multiple compilers on the same program, \toolname compares the behavioral equivalence of an original and its LLM translation, which requires handling cross-platform execution and floating-point tolerance differences inherent to heterogeneous HPC environments.
While each of these techniques has been applied individually in prior work, \toolname is the first to combine source-level metamorphic fuzzing, which generates diverse translation inputs to test LLM robustness, with runtime grammar-based fuzzing and differential testing, which detects semantic divergences in translated programs across diverse execution conditions.
\vspace{-1.80ex}
\section{Discussion}
\vspace{-0.8ex}
\label{sec:discussion}

\toolname's evaluation reveals a fundamental gap in how LLM-based HPC code translation is currently assessed.
Compilation success, token-level similarity, and developer-written tests from static benchmarks, the dominant evaluation proxies in prior work, substantially overestimate translation quality.
Programs that compile and pass developer-provided seed inputs can still contain latent semantic defects that manifest only under specific input conditions.
Developers relying on these proxies may deploy LLM-translated scientific code that produces incorrect results under production workloads.

The semantic errors identified in RQ3 share a common pattern: they involve subtle departures from the original execution semantics that preserve the program's structural appearance while altering its computational behavior. Intermediate variable elimination, execution model assumption transfer, loop bound errors, missing statement faults, shared memory scope mistranslation, and multi-dimensional index flattening all fall into this category. These errors are qualitatively different from compile-time errors in that they require behavioral testing to detect, and they are input-dependent in that they manifest only under specific runtime conditions. This reinforces the necessity of fuzzing-based evaluation for LLM-translated HPC code.

The 0\% compilation rate for full-program translation for $ChatPORT_{\text{CL\_13B}}$ highlights the limitations of kernel-level fine-tuning for end-to-end translation.
While many kernel-level compilation errors persist at the application level, full-program translation introduces additional challenges in host-device memory management, offloading constructs, multi-file dependencies, and kernel orchestration that are largely absent in isolated kernel translation.
Future research on LLM-based HPC code migration should therefore move beyond compilation-based evaluation and isolated kernel benchmarks, with fine-tuning efforts specifically targeting the HPC-specific error categories identified in our taxonomies, as these represent the primary barriers to reliable end-to-end CUDA-to-OpenMP translation.

\toolname is designed to be applicable beyond CUDA-to-OpenMP translation.
The metamorphic fuzzing component requires only a source language parser and a set of semantics-preserving mutation operators, both of which can be developed for any HPC programming model pair.
The differential testing component requires the ability to execute both the original and translated program on the same inputs and compare outputs, which is possible for any two programming models that can run on available hardware.
Extending \toolname to CUDA-to-SYCL, CUDA-to-Kokkos, and Fortran-to-C++ translation are natural directions for future work.

\vspace{-1.5ex}
\section{Threats to Validity}
\vspace{-0.8ex}
\label{sec:threats}

\textbf{Construct Validity.}
Our correctness metric flags a translation as incorrect if its output differs from the original by more than a configurable threshold on any fuzzed input. In our evaluation, correctness is determined by each application's built-in verification logic, which varies per application. When no built-in verification is available, \toolname uses a user-configured error norm threshold. This means correctness checking strictness is not uniform across applications, but reflects the domain-appropriate tolerance intended by the original developers.
A stricter threshold might flag benign floating-point variations as errors, while a looser threshold might miss genuine semantic divergences. We mitigate this by using three error norms (L1, L2, Max) and requiring exact equality for computations where floating-point variation is not expected.
The compile-time error taxonomy is qualitative and involves judgment in assigning errors to categories. To mitigate subjectivity, multiple authors independently categorized errors and reconciled disagreements, following established practices for qualitative analysis in software engineering research~\cite{Saldana2021}.

\textbf{Internal Validity.}
Fuzzing-based results depend on random seeds. We ran 5~independent fuzzing campaigns per application for source code mutation and 5~fuzzing campaigns per variant for differential testing to reduce sensitivity to individual seed choices. Three benchmark applications (\emph{chemv}, \emph{nbody}, \emph{axhelm}) were excluded from kernel-level evaluation due to ChatPORT's single-file limitation; this reflects a real constraint developers would face rather than a limitation of \toolname. The full-program translation prompt was engineered on 5~held-out examples using a consistent template across all LLMs, minimizing contamination risk.

\textbf{External Validity.}
Our evaluation uses 16~HeCBench applications spanning 7~scientific domains for runtime correctness evaluation and 47~applications for compile-time analysis. Very large applications with complex inter-kernel dependencies or non-standard memory management patterns are not represented.
Our findings are specific to CUDA-to-OpenMP translation using three ChatPORT variants; results may differ for other translation task or LLMs.

\vspace{-1.5ex}
\section{Conclusion}
\vspace{-0.8ex}
\label{sec:conclusion}

LLM-based translation of HPC programs is an increasingly important approach for porting scientific applications across heterogeneous programming models.
However, current evaluation practices, which rely on compilation success, token-level similarity, and developer-written tests from static benchmarks, cannot reliably ensure behavioral correctness.

We presented \toolname, a framework that combines metamorphic testing, grammar-based fuzzing, and differential testing to evaluate the behavioral correctness of LLM-translated HPC applications.
Evaluated on CUDA-to-OpenMP translation of 16~scientific applications from 7~domains using three top-performing ChatPORT variants at both kernel-level and full-program granularity, our evaluation produced four findings: neither compilation success nor developer-written tests can reliably ensure semantic correctness, as results range from perfect correctness to zero correctness across applications with similar compilability rates;
LLM-translated programs exhibit systematic compile-time error patterns across 9~kernel-level and 27~full-program categories;
semantic errors that survive compilation are application and input-dependent and require differential testing with diverse inputs to expose, with six categories identified; and
full-program translation is substantially harder than kernel-level translation, failing to compile entirely for $ChatPORT_{\text{CL\_13B}}$, and achieving at most 28\% compilability across the remaining two top-performing ChatPORT variants.

These findings establish that correctness-oriented evaluation is necessary for building trust in LLM-assisted HPC code porting. The error taxonomies provide actionable guidance for improving future LLM fine-tuning strategies, and extending \toolname to other programming model pairs such as CUDA-to-SYCL and CUDA-to-Kokkos is a natural direction for future work.

\bibliographystyle{ACM-Reference-Format}
\bibliography{references}

\end{document}